\documentstyle[aps,multicol,amsmath]{revtex}
\tighten
\newcommand{\bleq}{\ifpreprintsty \else
\end{multicols}\vspace*{-3.5ex}{\tiny \noindent\begin{tabular}[t]{c|}
\parbox{0.493\hsize}{~} \\ \hline \end{tabular}} \fi}
\newcommand{\eleq}{\ifpreprintsty \else
{\tiny\hspace*{\fill}\begin{tabular}[t]{|c}\hline
\parbox{0.49\hsize}{~} \\
\end{tabular}}\vspace*{-2.5ex}\begin{multicols}{2} \fi}
\newcommand{\bcols}{\ifpreprintsty\else\begin{multicols}{2}\fi}
\newcommand{\ecols}{\ifpreprintsty\else\end{multicols}\fi}

\newcommand{\non}{}
\newcommand{\be}{\begin{eqnset}\,}
\newcommand{\ee}[1]{#1\end{eqnset}}

\newcommand{\nn}{\global\let\non=\nonumber}

\newcommand{\sfrac}[2]{\mbox{$\frac{#1}{#2}$}}

\newcommand{\saverage}[1]{\langle#1\rangle}
\newcommand{\average}[1]{\left\langle#1\right\rangle}
\newcommand{\cumulant}[1]{\langle\!\langle#1\rangle\!\rangle}

\newcommand{\identity}{\mathrm{1}\hspace{-0.45ex}\mathrm{l}}
\newcommand{\eql}[1]{\global\let\non=\relax\label{eq:#1}%
\quad^{\underline{\{#1\}}}
}
\newcommand{\eq}[1]{(\ref{eq:#1})}

\newlength{\spacelen}
\newenvironment{eqnset}
{%       
 \global\let\non=\nonumber
  \vspace{-0.6ex}
 \begin{eqnarray}%
 \everymath{\displaystyle}%
 \begin{array}{[r@{\;\:}c@{\;\:}l]@{\vspace{1.6ex}}}%
}%
{%
 \vspace{-1.0ex}%
 \end{array}\non\end{eqnarray}\hspace{-\spacelen}%
}
\newlength{\ddone}\ddone 0pt
\newlength{\lzero}\lzero 0pt
\newcommand{\dd}[2]{%
        \settowidth{\ddone}{$#1$}%
        \ifdim\ddone=\lzero%
                \partial_{#2}%
        \else%
                {\frac{\partial #1}{\partial #2}}%
        \fi%
}
\newlength{\ddhone}\ddone 0pt
\newcommand{\ddh}[2]{%
        \settowidth{\ddhone}{$#1$}%
        \ifdim\ddhone=\lzero%
                (d/d#2)%
        \else%
                {\frac{d #1}{d #2}}%
        \fi%
}

\newcommand{\poisson}[1]{\{#1\}}
\newcommand{\fastterm}[1]{\mbox{ fast term in $#1$}}

\newcommand{\doubles}[1]{\left.#1\!\right.}

\newcommand{\ket}[1]{| #1 \rangle}
\newcommand{\bracket}[2]{\langle #1 | #2 \rangle}
\newcommand{\bbracket}[2]{\big\langle #1 \big| #2 \big\rangle}

\renewcommand{\vec}[1]{{\bf #1}}
\newcommand{\vek}{\vec{k}}
\newcommand{\veq}{\vec{q}}
\newcommand{\veqp}{\vec{q}'}
\renewcommand{\eql}[1]{\label{eq:#1}}
\renewcommand{\eq}[1]{Eq.~(\ref{eq:#1})}
\newcommand{\baverage}[1]{\bigl\langle #1 \bigr\rangle}

\newcommand{\Qarg}[2]{_{#2}^{#1}}
\newcommand{\QQarg}[4]{_{#2,#4}^{#1,#3}}
\newcommand{\QQQarg}[6]{_{#2,#4,#6}^{#1,#3,#5}}
\newcommand{\QQQargg}[6]{_{#2,#4,#6}^{#1#3#5}}

\newcommand{\QQzQargg}[6]{_{#2,#4;#6}^{#1#3#5}}

\newcommand{\inversion}{\mbox{\sf S}}
\newcommand{\barM}{{\bar M}}

\newcommand{\vecr}{{\vec r}}
\newcommand{\vecp}{{\vec p}}
\newcommand{\hatA}{{\hat A}}
\newcommand{\hatG}{{\hat G}}
\newcommand{\calA}{{\cal A}}
\newcommand{\calB}{{\cal B}}
\newcommand{\calC}{{\cal C}}
\newcommand{\calM}{{\cal M}}
\newcommand{\calP}{{\cal P}}
\newcommand{\calH}{{\cal H}}
\newcommand{\calL}{{\cal L}}
\newcommand{\calfastL}{{\cal L}_\perp}
\let\oldperp\perp
\renewcommand{\perp}{{\!\oldperp}}

\begin{document}

\title{Mode coupling theory for multi-point and multi-time correlation
functions}
\author{Ramses\ van Zon and Jeremy\ Schofield}
\address{Chemical Physics Theory Group, Department of Chemistry,\\
University of Toronto,Toronto, Ontario, Canada M5S 3H6}
\date{\today}
\maketitle

\begin{abstract}
We present a theoretical framework for higher-order correlation
functions involving multiple times and multiple points in a classical,
many-body system.  Such higher-order correlation functions have
attracted much interest recently in the context of various forms of
multi-dimensional spectroscopy, and have found an intriguing
application as proposed measures of dynamical heterogeneities in
structural glasses.  The theoretical formalism is based upon
projection operator techniques which are used to isolate the slow time
evolution of dynamical variables by expanding the slowly-evolving
component of arbitrary variables in an infinite, ``multi-linear''
basis composed of the products of slow variables of the system.  Using
the formalism, a formally exact mode coupling theory is derived for
multi-point and multi-time correlation functions.  The resulting
expressions for higher-order correlation functions are made tractable
by applying a rigorous perturbation scheme, called the $N$-ordering
method, which is exact for systems with finite correlation lengths in
the thermodynamic limit.  The theory is contrasted with standard mode
coupling theories in which the noise or fluctuating force appearing in
the generalized Langevin equation is assumed to be Gaussian, and it is
demonstrated that the non-Gaussian nature of the fluctuating forces
leads to important contributions to higher-order correlation
functions.  Finally, the higher-order correlation functions are
evaluated analytically for an ideal gas system for which it is shown
that the mode coupling theory is exact.
\vspace{1em}\\
PACS numbers: %
61.20.Lc, % Time-dependent properties; relaxation 
05.20.Jj, % classical fluids
05.40.-a. % fluctuation phenomena
%45.50.Jf,  Dynamics and kinematics: few and many body system
%47.10+g.  fluid dynamics general theory
%34.10.+x General theories and models of atomic and molecular
%collisions and interactions (including statistical theories,
%transition state, stochastic and trajectory models, etc.)
\end{abstract}

\bcols

\section{Introduction}

In a glassy system where structural frustration prevents relaxation to
equilibrium, dynamical properties often demonstrate complicated time
dependence\cite{Ediger}.  For instance, in dense colloidal systems, at
a given time some regions of the complex fluid are essentially static
and crystalline, while the dynamics in other regions exhibit behavior
which is associated with fluids.  In these systems, structural
rearrangement occurs through relatively rapid, collective, string-like
motions\cite{Kob,MSR99}. Furthermore, at later times, a region of the
fluid which previously appeared crystalline may exhibit fluid-like
properties.  Such heterogeneous behavior is characteristic for
structural glasses and super-cooled complex fluids
\cite{Li89,Spiess91,Moynihan93,Cicerone95,Schiener96,Reichert97}.  To
describe this behavior, it is natural to examine how the local density
of the liquid is correlated over various spatial
domains\cite{Dasgupta,Glotzer}, or, when one is more interested in the
different time scales of slow global changes of structure and the
local decay of correlations, multiple time correlation
functions\cite{Heueretal}. Both types of higher-order correlation
functions have recently been proposed and used as measures for
characterizing ``dynamical heterogeneities'' in structural glasses.

A number of experimental probes of examining detailed dynamical
features taking place on various length and time scales in glassy
systems have emerged over the last few years.  These new approaches
have the potential to provide extremely useful information on how
collective motions of the system are correlated to specific
statistical features of the dynamics such as the distribution of time
scales of fluctuations, the length scale and size-distribution of
solid-like clusters, and cage structural relaxation rates.  One
approach to probe the nature of dynamical heterogeneities is based on
single molecule spectroscopy techniques
\cite{Moerner94,Lu97,Basche95,VandenBout2001} in which the environment
of a one molecule is tracked over a period of time.  The technique
allows the information of not only the distribution of heterogeneous
environments but also the explicit reorganization times which are
present in the system, since the individual measurements are not
statistically averaged.  A somewhat different experimental approach is
based on multi-dimensional NMR\cite{Sillescu96,Spiess98} and
non-resonant non-linear Raman spectroscopy
\cite{Fleming,Miller,Tokmakoff}.  The response function in these
experiments can be related to higher-order correlation functions using
response theory\cite{Tanimura97,Ohmine98}.

Given the interest in multi-time and multi-point correlation
functions, the need for a theory that accurately predicts these
quantities is clear. Surprisingly, there has been relatively little
work along these lines and the literature is not nearly as extensive
as it is for ordinary, two-time, two-point correlation functions such
as the dynamic structure factor.  Although there have been several
recent microscopic theories for the off-resonant fifth-order response
function for simple liquids\cite{Stratt,Reichman}, little work has
been done on constructing microscopic theories for general
higher-order correlation function since the kinetic theories of De
Schepper and coworkers\cite{dSE74}, who attempted to extract the
non-analytic density contributions to the Burnett coefficients in hard
sphere liquids.

A common approach to describing the dynamical properties of liquids
and complex fluids at long times is based on the generalized Langevin
equation \cite{Hansen}.  The basic utility of the generalized Langevin
equation depends on the assumption that the long-time behavior of an
arbitrary dynamical variable of the system can be written in terms of
the dynamics of a specific set of slow modes.  This slow behavior can
be isolated by extracting the projection of the dynamical variable
onto the slow modes, which effectively form a basis set for the long
time behavior of the system.  This approach has been successfully
applied to describe relaxation and simple time correlation functions
in a wide variety of condensed phase systems.  In the context of
simple liquid systems, it was initially assumed that only the linear
densities of conserved variables of the system composed the set of
slow variables of the system \cite{Schofield66,Desai71,Murase70}.
However, it was discovered that this basis set was insufficient
\cite{Michaels75} to describe the complex, non-exponential relaxation
of simple time correlation functions observed in molecular dynamics
simulations \cite{Alder68}.  The subsequent observation that theories
incorporating multi-linear products of the linear densities in the
basis were capable of yielding the correct asymptotic long-time
behavior of simple correlations \cite{Ernst72,Pomeau72} led to the
development of ``mode-coupling'' theories.  The seminal work of
Kawasaki \cite{K70}, who proposed that the linear Langevin equation be
replaced by a non-linear version in which the fluctuating forces obey
Gaussian statistics, sparked the later development of kinetic
mode-coupling theory models of dense liquids
\cite{Gotze75,Munakata77,Bosse78,Gaskell78,Sjogren79}.  At roughly the
same time, Ronis\cite{R81} used the framework of Kawasaki model
coupling theory to formulate a theory of higher-order correlation
functions in which the multi-linear slow variables forming the basis
set for long-time evolution in the system obey Gaussian statistics.
Although Ronis' theory contains a number of assumptions, it is the
assumption of Gaussian statistics which leads to clear inconsistencies
in the predictions for higher-order correlation functions.
Unfortunately, the Gaussian assumption is fundamental in the Kawasaki
formulation of mode coupling theory, and is difficult to generalize.

The purpose of this paper is to provide a solid theoretical framework
to calculate multi-point and multi-time correlation functions. It is
based on the mode coupling theory obtained by a projection operator
formalism \cite{Zwanzig61,Mori65} as developed by Oppenheim and
coworkers \cite{MO82,SO81,SLO92,SO92}. By careful consideration of how
to consistently identify fast and slow behavior in time correlation
functions, we derive expressions for multi-point and multi-time
correlation functions in terms of simple time correlation functions.
Since the basis set for the slow modes is infinite, an infinite number
of terms arise in the expressions for the higher order correlation
functions.  It is demonstrated that if the system has a finite
correlation length, the infinite series can be truncated by applying
the so-called $N$-ordering perturbation expansion method,  which is 
exact in the thermodynamic limit. The use of
this perturbation scheme circumvents the need to assume that the basis
set obeys Gaussian statistics.
Based on this method, the leading order expressions and first
mode-coupling corrections for higher order correlation functions are
presented.  Finally, it is shown how the formalism applies for an
ideal gas system for which it is demonstrated that the theory yields
the exact result for simple, multi-point, and multi-time correlation
functions.

\section{Two-time correlations}

\subsection{The system and slow variables}

Consider a classical system composed of $N$ point particles in which
the momentum and position of particle $i$ are denoted by $\vec p_i$
and $\vec r_i$, respectively. Given the Hamiltonian $\calH$, a
function $B(\Gamma)$ of the phase point $\Gamma=(\vec r_1, \cdots,
\vec r_N , \vec p_1, \cdots, \vec p_N )$ evolves according to
\[
	\dot B(t) = \poisson{B(t),{\calH}} \equiv {\calL} B(t),
\]
where $\{,\}$ denotes the Poisson bracket and ${\calL}$ is the
Liouville operator for the system.  The evolution equation can be
solved formally as $B(t) = \exp(\calL t)B(0)$, where here and below
$B(t)$ is taken to denote $B(\Gamma (t) )$.

Typically, each dynamical variable of the system can be separated into
slowly-varying and quickly-varying parts.  We will assume that the
time-dependent correlation functions of the quickly-varying components
decay to zero on a short (called ``microscopic'') time scale
$\tau_{m}$, whereas correlation functions of the slowly varying part
decay on a longer time scale $\tau_{h}$.  Hence, at long times
%$\tau_{h} \ll t \ll \tau_{m}$,  
$t \gg \tau_{m}$, the decay of an arbitrary correlation function can
be described by the decay of its slow component.  In what follows, we
postulate that the slowly varying part of an arbitrary dynamical
variable is an analytic function of a set of slow variables of the
system.  In this sense, the slow variables form a basis set in which
to represent the long-time behavior of the system.

To identify slow variables in such a system, one considers all the
conserved quantities, which can be taken together in one column vector
$A$, with components $A^a$. When these quantities are extensive, they
can be expressed as a sum over contributions from the individual
particles,
\begin{equation}
	A(\Gamma) = \sum_{j=1}^N a_j(\Gamma) ,
\eql{1}
\end{equation}
which leads to the identification of the densities as the local
version of $A$,
\begin{equation}
	A(\vec r;t) = \sum_j a_j(t) \delta(\vec r-\vec r_j(t)),
\eql{2}
\end{equation}
with Fourier components
\begin{equation}
	A_{{\vec{k}}}(t) = \sum_j a_j(t) e^{i{\vec{k}}\cdot\vec r_j(t)}.
\eql{3}
\end{equation}

For the case of a simple fluid of $N$ point particles, the extensive
slow variables are the number density, momentum density, angular
momentum density and the energy density, and in \eq{2}, one would use
the microscopic expressions for those quantities. For point particles,
the angular momentum density can be expressed in terms of the momentum
density and need not be included in $A$.
%For instance, if the Hamiltonian is given as
%\[
%	\calH = \sum_{i=1}^N \left [ \frac{|\vec p_i|^2}{2m} +
%		\sum_{j=1}^{i-1} \phi(\vec r_i-\vec r_j)\right],
%\]
%the energy can be written as $E=\sum_i e_i$ with
%\[  
%	e_i = \frac{|\vec p_i|^2}{2m}
%		+ \frac12 \sum_{j\neq i} \phi(\vec r_i-\vec r_j).
%\]
%So we see that $a_i$ in \eq{1}-\eq{3} doesn't necessarily depend on
%coordinates of particle $i$ only.

The small wave-vector components $A_{\vec{k}}$ of the densities
correspond to large length scale fluctuations, and these are expected
to evolve slowly since their time derivatives are proportional to
$|\vec{k}|$.  On the other hand, large wave-vectors with magnitudes
beyond some cut-off value $k_{c}$ correspond to small length-scale
fluctuations and have large time derivatives provided the system is
not too dense. Thus, one can identify the slow variables to be
composed of the Fourier-transform of densities of conserved variables
whose wave-vector arguments $\vec{k}_{i}$ are restricted have
magnitudes less than $k_{c}$.

\subsection{Generalized Langevin equation}

The distinction between fast and slow behavior can be made more
precise on the level of correlation functions, following the
projection operator method as used by Oppenheim and
others\cite{MO82,SO81,SLO92,SO92}.

To isolate the slow part of a correlation function, one assumes it to
be an analytic function of the slow variables $A_{\vec{k}}$. The
non-linear dependence of the time correlation functions on the slow
variables can be incorporated within the framework of projection
techniques by employing a basis of non-linear functions of
$A_{\vec{k}}$, the so-called multi-linear basis.

Using the ensemble average $\saverage{\cdots}$ as an inner product, we
define a projection operator $\calP_1$ that projects onto the
deviations $\hatA_{\vec{k}}\equiv A_{\vec{k}} -
\saverage{A_{\vec{k}}}$ of the slow variables from their equilibrium
value, as
\begin{equation}
 \calP_1 X \equiv \saverage{X\hatA_{\vec{k}}} * K^{-1}_{\vec{k}}
		  *\hatA_{\vec{k}},
\eql{P1}
\end{equation}
where 
\[
	K\Qarg{}{\vec{k}} = \saverage{\hatA_{\vec{k}}\hatA_{\vec{k}}^*}
\]
denotes the normalization.  The ``$*$''-product in \eq{P1} includes a
sum over components of the column vector $A_{\vec{k}}$ -- the {\em
hydrodynamic indices} -- as well as over wave-vectors ${\vec{k}}$. For
a translationally invariant system, all wave-vectors in an average
must add to zero, so only when ${\vec{k}}$ equals the sum of
wave-vectors in $X$ will there be a contribution in \eq{P1}.

Using only the linear projection $\calP_1$ will not give a proper mode
coupling theory since the long-time dependence of dynamical variables
due to multi-linear orders of the densities is not extracted by the
projection operator\cite{SLO92,SO92,K77,KO73}. It is therefore
convenient to define an orthogonal, multi-linear basis as
\begin{eqnarray}
	Q_0 &\equiv &1
\nonumber\\ 
	Q\Qarg{a}{\vec k} &\equiv& A^a_{\vec k} - \calP_0 A^a_{\vec k}
		\equiv \hatA^a_{\vec k}
\nonumber\\ 
	Q\QQarg{a}{\vek-\veq}{b}{\veq} &\equiv&
	\hatA^a_{\vek-\veq}\hatA^b_{\veq} 
	- (\calP_0+\calP_1)\hatA^a_{\vek-\veq}\hatA^b_{\veq}
\eql{Qset}
\\ &&\vdots
\nonumber
\\
	Q\QQQarg{a_1}{{\vec{k}_1}-{\vec{k}'}}
	        {a_2}{{\vec{k}}_2}
	        {\ldots a_{n}}{\ldots ,{\vec{k}}_{n}}
	&\equiv &
	\bigg[{\identity} - \sum_{j=0}^{n-1}\calP_j\bigg]
	\hatA^{a_1}_{{\vec{k}_1}-{\vec k'}}\hatA^{a_2}_{{\vec{k}}_2}
	\cdots\hatA^{a_n}_{{\vec{k}}_{n}}, 
\nonumber
\end{eqnarray}
where ${\vec{k}'}=\sum_{l=2}^{n}{\vec{k}}_l$, and the projection
operators $\calP_j$ are defined as
\begin{equation}
  \calP_jX = \sum_{|\alpha|=j}\saverage{XQ_{\hat\alpha}}
		*K^{-1}_{\hat\alpha\alpha}*Q_\alpha,
\eql{Pi}
\end{equation}
In this notation, a Greek index denotes a set of wave-vectors and
hydrodynamic indices, and $|\alpha|$ denotes the number, or {\it mode
order}, of hydrodynamic indices in a set $\alpha$.  For simplicity of
notation, the full notation will often be written $Q_1$, $Q_2$, $Q_3$,
etc., for $Q_\alpha$ when $|\alpha|=1$, $2$, $3$, $\ldots$.
Furthermore, non-hatted and hatted Greek indices will always have the
same mode order, i.e.  $|\alpha|=|\hat\alpha|$, but represent
different sets of wave-vectors and/or hydrodynamic indices. Finally,
in \eq{Pi} $K_{\hat\alpha\alpha}=\saverage{Q_{\hat\alpha}Q_\alpha}$,
and the ``$*$''-product now includes a sum over all hydrodynamic
indices and wave-vectors in the summation indices ($\alpha$ and
$\hat\alpha$). To make sure the contributions from the same component
in a sum over $\alpha$ are not over-counted, one has to divide by the
number of ways the indices can be rearranged in $\alpha$.

The projection terms in \eq{Qset} force the set $Q_\alpha$ to be
orthogonal in mode order so that $\saverage{Q_\alpha Q_\beta}=0$
unless $|\alpha|=|\beta|$.  This property will be not only convenient
but very important in the subsequent analysis.  By assumption, the
long-time behavior of an arbitrary variable $C$ can be isolated by the
projection operator,
\begin{equation}
	\calP C \equiv \sum_{j=0}^{\infty} \calP_jC = \saverage{C Q^*_\alpha} 
	K^{-1}_{\alpha\hat\alpha}  Q_{\hat\alpha},
\end{equation}
where we have used the convention that repeated Greek indices imply a
``$*$''-product and a summation over mode-order.  This notation will
be used throughout this article unless stated otherwise.

Applying operator identity
\[
	e^{At} = e^{(A+B)t} - \int_0^t e^{A(t-\tau)} B e^{(A+B)\tau} d\tau,
\]
to the evolution equation, one easily obtains,
\begin{equation}
\begin{split}
	e^{\calL t} = &
	e^{\calL t}\calP 
	+ e^{\calfastL t}\calP_\perp
\\
	&+\int_0^t e^{\calL(t-\tau)} \calP\calL 
	e^{\calfastL\tau} \calP_\perp \,d\tau,
\end{split}
\eql{eLtexp}
\end{equation}
where $\calP_\perp = 1-\calP$ and $\calfastL=\calP_\perp\calL$.  We
apply this operator to $\dot{Q}_\alpha$, and write out $\calP$ to get
\begin{eqnarray}
	\dot{Q}_\alpha(t) &=& 
	\saverage{\dot{Q}_\alpha Q^*_\beta} 
	K^{-1}_{\beta\hat\beta}  Q_{\hat\beta}(t) 
	+ \phi_\alpha(t)
\nonumber\\
&&
	+ \int_0^{t}
		e^{\calL(t-\tau)}\saverage{
		Q^*_\beta
		\calL
		e^{\calfastL\tau}\calP_\perp\,
		\dot{Q}_\alpha }
		 K^{-1}_{\beta\hat\beta}  Q_{\hat\beta}
	d\tau
\nonumber
\\
&
	= &\int_0^t M_{\alpha\beta}(\tau) 
	Q_\beta(t-\tau) \,d\tau + \phi_\alpha(t),
\eql{qdottwo}
\end{eqnarray}
which is the generalized Langevin equation, where
\begin{equation}
	M_{\alpha\beta}(\tau) \equiv \big[ 2\delta(\tau)
	\saverage{\dot{Q}_\alpha Q^*_{\hat\beta}}
	-\saverage{\phi_\alpha(\tau)\phi^*_{\hat\beta}}\big]
	 K^{-1}_{\hat\beta\beta},
\eql{M}
\end{equation}
and the fluctuating force $\phi_\alpha(t)$ is defined by
\begin{equation}
	\phi_\alpha(t) \equiv 
e^{ \calfastL t } \calfastL Q_\alpha
=e^{\calfastL t}\calP_\perp \dot{Q}_\alpha
\eql{flucforce}
.
\end{equation}

Under the assumption that $\calP$ projects out all the slow behavior,
the matrix $M_{\alpha\beta}(\tau)$ is ``fast'' in the time variable
$\tau$ in the sense that, for a microscopic time scale $\tau_m$ much
smaller then the hydrodynamic time scale $\tau_h$ of the slow
variables,
\begin{equation}
	\saverage{\phi_\alpha(\tau) B} \approx 
	\int_0^\infty\saverage{\phi_\alpha(t) B} dt\, \delta(\tau) + O(\tau_m/\tau_h).
\eql{fastness1}
\end{equation}
It then follows that the main contribution in the integral in
\eq{qdottwo} comes from small $\tau$, and we can approximate
\eq{qdottwo} by one which is local in time,
\begin{equation}
	\dot{Q}_\alpha(t) = \bar{M}_{\alpha\beta} Q_\beta(t) +
\phi_\alpha(t), \eql{qdotthree}
\end{equation}
where the instantaneous matrix is given by
\begin{equation}
	\bar{M}_{\alpha\beta} =  \int_0^\infty M_{\alpha\beta}(t)\, dt
\eql{instantM}.
\end{equation}

It will become clear in section \ref{sec:renorm} that the assumption
of a fast decaying memory kernel in \eq{qdottwo} is not in
contradiction with the slow memory kernel that occurs in {\em linear}
mode coupling equations like those used in the study of the glass
transition\cite{Gotze75}.

Defining the time correlation functions of the basis set to be
\begin{eqnarray}
	G_{\alpha\beta}(t) &\equiv& \saverage{Q_\alpha(t)Q_{\hat\beta}^*} 
	 K^{-1}_{\hat\beta\beta} ,
\eql{Gdef}
\end{eqnarray}
and using \eq{qdottwo}, one obtains the simple expression for the time
correlation function,
\begin{eqnarray}
	\dot{G}_{\alpha\beta}(t) = 
	\int_0^t M_{\alpha\delta}(\tau) G_{\delta\beta}(t-\tau)\,d\tau
,
\eql{Gdot}
\end{eqnarray}
where we have used the fact that
$\saverage{\phi_{\alpha}(t)Q_{\beta}^{*}}=0$ by construction.

In the instantaneous approximation [corresponding to \eq{fastness1}],
\eq{qdotthree} yields $\dot{G}_{\alpha\beta}(t)=\barM_{\alpha\delta}
G_{\delta\beta}(t)$, which can be integrated to obtain
\begin{equation}
	G_{\alpha\beta}(t) = [e^{\barM t}]_{\alpha\beta}.
\eql{instantG}
\end{equation}
If we do not wish to make the instantaneous approximation, it is
easiest to look at the Laplace transform of the time correlation
function
\[
	G_{\alpha\beta}(z) \equiv 
	\int_0^\infty G_{\alpha\beta}(t) e^{-zt} \,dt .
\]
Since the time convolution in \eq{Gdot} is a simple product in Laplace
space, we have
\begin{equation}
	G_{\alpha\beta}(z) = \saverage{Q_\alpha(z)Q^*_{\hat\beta}} 
	K^{-1}_{\hat\beta\beta} = 
	\big[z{\identity}-M(z)\big]^{-1}_{\alpha\beta},
\eql{14b}
\end{equation}
where the $\alpha\beta$ element of the inverse is meant, and
$M_{\alpha\beta}(z) = [\saverage{\dot{Q}_\alpha Q^*_{\hat\beta}}
-\saverage{\phi_\alpha(z)\phi^*_{\hat\beta}}]
K^{-1}_{\hat\beta\beta}$.

Note that one is frequently interested in just the part of
$G_{\alpha\beta}(t)$ that involves the linear variables, corresponding
to such readily observable physical quantities as the time correlation
functions of the linear density, momentum and energy.  These are given
by $G_{11}(t)$, the $|\alpha| = |\beta| = 1$ sub-block of the
infinite-dimensional matrix $G_{\alpha\beta}(t)$.

In the hydrodynamic limit in which the magnitude of the wave-vectors
$\vec{k}$ is small, the wave-vector can be used to perturbatively
order various contributions to $M_{\alpha\beta}(z)$.  Noting that each
time derivative in \eq{M} brings down a factor of ${\vec{k}}$, the
first term in \eq{M}, called the {\em Euler} term, is of order
${\vec{k}}$, whereas the second, or {\em dissipative} term, is
$O(|{\vec{k}}|^2)$. On the basis of these arguments, one might think
that the dissipative term in $M_{\alpha\beta}(z)$ can be neglected in
the hydrodynamic limit.  However, examining the form of the densities,
it is easy to see that in fact the Euler term is imaginary and can
only give rise to oscillatory behavior\footnote{The Euler term can
give rise to Gaussian relaxation if one has an infinite number of slow
modes, see Sec.~\ref{sec:ideal}.}. It is therefore clear that the
dissipative term must be included in order to obtain solutions to
\eq{qdotthree} which are well-behaved in the long-time limit.

\subsection{$N$-ordering}

%The results of the previous section are formally exact but not very
%useful because an infinite number of multi-linear modes is required to
%evaluate even simple time correlations.  Note that
%two-time correlation functions at a particular mode order are given
%by specific sub-blocks of the inverse of an infinite-dimensional
%matrix.  
Since two-time correlation functions at a particular mode-order are
given by specific blocks of the inverse of an infinite dimensional
matrix, even the calculation of simple time correlation functions
requires an infinite number of multi-linear modes. Thus, the results
of the previous section are formally exact, but not very useful.

The evaluation of the time correlation functions is greatly
facilitated by applying cumulant expansion or $N$-ordering techniques.
The $N$-ordering scheme was first introduced by Machta and Oppenheim
\cite{MO82} as an extension of Van Kampen's inverse system size
expansion ($\Omega$ expansion)\cite{VanKampen} and developed further
in Ref.~\cite{SLO92}.  It is essential in obtaining the correct
Stokes-Einstein law for a Brownian particle using mode coupling
techniques \cite{SO92}.

In the $N$-ordering approach, one assigns to each cumulant of an
average appearing in the equations an order of $N$ (the number of
particles). The starting point is to consider cumulants, which we
denote by ``$\cumulant{ \cdots }$''.  For a product of linear
densities, the cumulant expansion consists of all possible ways of
combining the densities into groups, i.e.,
%\begin{eqnarray*}
%   \average{A}&=&\cumulant{A}
%\\
%   \average{AB}&=&\cumulant{AB}+\cumulant{A}\cumulant{B}
%\\
%   \average{ABC}&=&\cumulant{ABC}
%	+ \cumulant{AB}\cumulant{C}
%\\&&
%	+ \cumulant{AC}\cumulant{B}
%	+ \cumulant{BC}\cumulant{A}
%\\&&
%	+ \cumulant{A}\cumulant{B}\cumulant{C},
%\end{eqnarray*}
$\average{A}=\cumulant{A}$,
$\average{AB}=\cumulant{AB}+\cumulant{A}\cumulant{B}$, and so on. The
assignment of $N$-orders is based on the observation that each
cumulant containing $n$ linear densities is of order
$N(\xi/a)^{3(n-1)}$, where $a$ is the average distance between
particles\cite{MO82}. Hence, the requirements for the expansion method
to be meaningful are that the system should have a finite [$O(N^0)$]
correlation length $\xi$, and that the integrated densities should be
extensive. For an extension to non-extensive quantities like tagged
particle densities, see Ref.~\cite{SO92}.

For instance, the cumulant of a linear-linear correlation function of
the number density $N_{\vec{k}}$, defined as $\sum_{j=1}^N
e^{i{\vec{k}}\cdot\vec r_j}$, can be formally written as
\begin{eqnarray*}
	\cumulant{N_{{\vec{k}}}N_{{\vec{k}}}^*}&=& 
	\saverage{N_{{\vec{k}}}N_{{\vec{k}}}^*} 
	-\saverage{N_{{\vec{k}}}}\saverage{N_{{\vec{k}}}^*}\delta_{k0}
\\&
	=& [\saverage{N} + \saverage{N(N-1)e^{i{{\vec{k}}}\cdot({\vec
	r}_1-{\vec r}_2)}}] (1-\delta_{{\vec{k}} 0}) \\&& +
	[\saverage{N^2}-\saverage{N}^2]\delta_{{\vec{k}} 0}.
\end{eqnarray*}
For ${\vec{k}}=0$, the cumulant expansion of
$\saverage{N_{{\vec{k}}}N_{{\vec{k}}}^*}$ is given by
$\saverage{(N-\saverage{N})^2}$, which, in the grand canonical
ensemble, is of order $\saverage{N}$. For ${\vec{k}}\neq 0$, the
expression is proportional to $N$ as well, because particles beyond
the correlation length $\xi$ will not contribute to the average of
$\exp(i{\vec{k}}\cdot({\vec r}_1-{\vec r}_2))$, so that
\[
	\saverage{N(N-1)\exp(i{\vec{k}}\cdot({\vec r}_1-{\vec r}_2))}
	\propto N (N-1)\frac{\xi^3}{V} = O(N).
\]

The $N$-ordering of various quantities has been discussed extensively
elsewhere\cite{MO82,SLO92,SO92}, so here we just state a few
properties of the procedure that enable us to estimate $N$-orders of
relevant quantities.

One important basic property of averages involving $Q_\alpha$ is that
when one has an average of the form $\saverage{Q_\alpha(t) B Q^*_1}$,
where $B$ can be a product of $Q_1$ again, it can be factored into a
term of order $N^{2}$,
\begin{equation}
 \saverage{Q_\alpha(t) B Q\Qarg{a*}{{\vec{k}}}} \approx
	\saverage{Q_{\alpha-1}(t)B}
	\saverage{Q\Qarg{a_j}{{\vec{k}}_j}(t)
	        Q\Qarg{a*}{{\vec{k}}}}
	\delta_{{\vec{k}}_j{\vec{k}}}
\eql{factorization}
\end{equation}
plus terms involving lower powers of $N$.  That is, the leading
$N$-order term in the cumulant expansion can be found by equating any
wave-vector ${\vec{k}}_j$ from the set $\alpha$ with ${\vec{k}}$. In
\eq{factorization}, $\alpha-1$ denotes the set $\alpha$ with
${\vec{k}}_j$ and $a_j$ removed. Hence, ``$\approx$'' means that the
expression is correct up to higher orders of $N$ and only for the case
where appropriate wave-vectors have been equated.  We can reduce such
an expression further by equating wave-vectors in $B$ with those in
the set $\alpha-1$. An important point, established in
Ref.~\cite{SLO92}, is that equating wave-vectors within the set
$\alpha$ does {\em not} increase the $N$ order due to the subtraction
terms in their definition \eq{Qset}.  Thus the orthogonalization
procedure used to construct the multi-linear basis plays an important
role in the proper $N$-ordering of correlation functions of multi-linear
densities.

Using the ordering scheme, one obtains
$K_{\alpha\hat{\alpha}}=O(N^j)$, where $j$ is the number of matched
sets of wave-vectors in $\alpha$ and $\hat\alpha$, and hence the
$N$-ordering of its inverse is
\begin{equation}
	K^{-1}_{\alpha\hat\alpha} = O(N^{-|\alpha|}),
\eql{NorderKinv}
\end{equation}
irrespective of matching of wave-vectors.  In Ref.\cite{MO82}, it was
shown that
\[
	M_{\alpha\beta}(t) = \left\{\!\!\begin{array}{ll}
		O(1) &\mbox{if $|\alpha|\geq|\beta|$} \\
		O(N^{|\alpha|-|\beta|})&   \mbox{if $|\alpha|<|\beta|$}
	\end{array}\right. ,
\]
with the same $N$-ordering holding for $\barM_{\alpha\beta}$.  A
similar result can be obtained for $G_{\alpha\beta}(t)$
\begin{equation}
	G_{\alpha\beta}(t) = \left\{\!\!\begin{array}{ll}
		O(1) &\mbox{if $|\alpha|\geq|\beta|$} \\
		O(N^{|\alpha|-|\beta|})&   \mbox{if $|\alpha|<|\beta|$}
	\end{array}\right.
	.
\eql{N-ordering1}
\end{equation}

\subsection{Multi-point correlations}
\label{subsec:multipoint}

We define a multi-point correlation function to be any
$G_{\alpha\beta}(t)$ for which $|\alpha|$ or $|\beta|$ is greater than
1. The multi-point correlations functions therefore contain several
factors of the linear densities $A_{\vec{k}}$, and involve more than
one wave-vector.  Hence, in position space, the multi-point
correlation functions involve at least three spatial, or {\it two
relative}, position coordinates, justifying the name multi-point
correlation functions.

As $M_{\alpha\beta}(z)$ is $O(1)$ if $\alpha$ and $\beta$ have all
wave-vectors matched, and of lower order otherwise, the leading
$N$-order contributions to $G_{\alpha\beta}(t)$ can be obtained by
expanding \eq{14b} in the wave-vector-diagonal component of
$M_{\alpha\beta}(z)$, denoted by
$M^d_{\alpha\beta}=M_{\alpha\alpha'}(z)\delta_{\alpha'\beta}$. Here
primed Greek indices will always have the same set of wave-vectors as
as their unprimed variant, but not necessarily the same hydrodynamic
indices.  Defining the off-diagonal part as $M^o_{\alpha\beta}(z) =
M_{\alpha\beta}(z) - M_{\alpha\alpha'}(z) \delta_{\alpha'\beta}$, we
get from \eq{14b}
\begin{eqnarray}
	G_{\alpha\beta}(z) &=&
	\big[z{\identity}-M^d(z)-M^o(z)\big]^{-1}_{\alpha\beta}
\nonumber\\ &=& 
	\big[\{z{\identity}-M^d(z)\}\{{\identity} - \hatG(z)
	M^o(z)\}\big]^{-1}_{\alpha\beta}
\nonumber\\ &=& 
	\big[{\identity} - \hatG(z)
	M^o(z)\big]^{-1}_{\alpha\beta'} \hatG_{\beta'\beta}(z),
\eql{16b}
\end{eqnarray}
where we have defined  the diagonal in wave-vector
\begin{equation} 
	\hatG_{\alpha\alpha'}(z) =
	\big[z{\identity}-M^d(z)\big]^{-1}_{\alpha\alpha'}.
\eql{Gdadef}
\end{equation}
We can therefore expand \eq{16b} as
\begin{eqnarray}
	G_{\alpha\beta}(z) &=& \hatG_{\alpha\alpha'}(z)\delta_{\alpha'\beta}
	+\hatG_{\alpha\alpha'}(z) M^o_{\alpha'\beta'}(z)
	\hatG_{\beta'\beta}(z)
\nonumber\\&&
	+ \hatG_{\alpha\alpha'}(z) M^o_{\alpha'\gamma}(z)
	\hatG_{\gamma\gamma'}(z) M^o_{\gamma'\beta'}(z)
	\hatG_{\beta'\beta}(z) 
\nonumber\\&&+ \ldots
\eql{expansion}
\end{eqnarray}

Note that when a term is of lower order in $N$, this does not mean it
can be neglected in the thermodynamic limit.  The leading $N$-order is
found from equating wave-vectors, whereas the next order in $N$ often
comes with an un-restricted summation over wave-vectors.  If $M$
denotes the number of (slow) wave-vectors in the system, these
correction terms are $O(M/N)$. The number of wave-vectors $M$ grows
with the system size. Therefore such mode coupling corrections survive
in the thermodynamics limit. But when the mode coupling parameter
$M/N$ is small, as is typically the case for simple liquids at
moderate densities, it can be treated as perturbation
parameter\cite{SLO92}.

In \eq{expansion}, note the appearance of $\hatG_{\alpha\alpha'}$,
defined in \eq{Gdadef}.  Although this is diagonal in wave-vector, it
is {\em not} the diagonal part of $G_{\alpha\beta}$, as the terms
following the first term in \eq{expansion} give contributions for
$\alpha$ and $\beta$ diagonal. As was shown in Ref.~\cite{SLO92}, in
the thermodynamic limit, the diagonal part of $G_{\alpha\beta}$,
denoted by $G_{\alpha\alpha'}$, can be factored as
\begin{equation}
	G_{\alpha\alpha'}(t) \approx \sum_\sigma\Biggl[
	\prod_j
	G\Qarg{a_ja'_j}{{\vec{k}}_j}(t),
	\delta_{\vec{k}_j\vec{k}'_{\sigma_j}}
	\Biggr]
\eql{Galphafactor}
\end{equation}
where $a_j$ and $a'_{\sigma_j}$ are hydrodynamic indices from $\alpha$
and $\alpha'$, and $\vec{k}_j$ and $\vec{k}'_{\sigma_j}$ the
respective wave-vectors. The summation is over all permutations
$\sigma$ of the indices in $\alpha'$.  This factorization is obtained
also by cumulant expansion under the assumption that there is a finite
time dependent correlation length.

Thus, if we were able to express \eq{expansion} in terms of
$G_{\alpha\alpha'}$ instead of $\hatG_{\alpha\alpha'}$, we could
combine that with \eq{Galphafactor} to get an expression for any
multi-point function in terms of in the two-point correlation
functions $G_{11}$ and vertices $M_{\alpha\beta}^{o}$.  \eq{expansion}
can be re-expressed in this desired form by the following re-summation
of terms: We write $G_{\alpha\beta}=
G_{\alpha\alpha'}\delta_{\alpha'\beta} +G_{\alpha\beta}^o$, and use
the Dyson form of \eq{expansion}:
\[
G_{\alpha\beta} = \hat{G}_{\alpha\alpha'}\delta_{\alpha'\beta} +
			G_{\alpha\gamma}
			M^o_{\gamma\beta'}\hat{G}_{\beta'\beta}
\]
for the off-diagonal part to get
\[
	G_{\alpha\beta} = G_{\alpha\alpha'}\delta_{\alpha'\beta} +
			G_{\alpha\gamma} M^o_{\gamma\beta'}\hat
			G_{\beta'\beta}^{(\alpha)},
\]
where the superscript $(\alpha)$ means that $\beta$ is restricted to
not have the same wave-vector set as $\alpha$. Iterating this equation
yields
\begin{eqnarray*}
	G_{\alpha\beta} &=& G_{\alpha\alpha'}\delta_{\alpha'\beta} +
			G_{\alpha\alpha'} M^o_{\alpha'\beta'}
			\hat{G}_{\beta'\beta}^{(\alpha)}
\\&&
+			G_{\alpha\alpha'} 
			M^o_{\alpha'\gamma'}
			\hat{G}_{\gamma'\gamma}^{(\alpha)}
			M^o_{\gamma\beta'}
			\hat{G}_{\beta'\beta}^{(\alpha)}
	+ \ldots
\\
	&\equiv &
	G_{\alpha\alpha'}\delta_{\alpha'\beta}
	+
	G_{\alpha\alpha'}M_{\alpha\delta}^o\widetilde{G}^{(\alpha)}_{\delta\beta} ,
\end{eqnarray*}
where by definition, $\widetilde{G}_{\delta\beta}^{(\alpha)}$ has the
same form as the right-hand side of \eq{expansion}, with $\alpha$
replaced by $\delta$ and the
restriction that none of the wave-vector sets in the expression are
equal to $\alpha$. Being of that form, we can repeat this procedure
for $\widetilde{G}_{\delta\beta}^{(\alpha)}$ to obtain,
\begin{eqnarray*}
	G_{\alpha\beta} &=&
	G_{\alpha\alpha'}\delta_{\alpha'\beta}
	+
	G_{\alpha\alpha'}M_{\alpha\delta}^o
	\widetilde{G}^{(\alpha)}_{\delta\beta}
	\\
	& & +
	G_{\alpha\alpha'}M_{\alpha\gamma}^o
	\widetilde{G}^{(\alpha)}_{\gamma\gamma'}
	M_{\gamma\delta}^o
	\widetilde{G}^{(\alpha,\gamma)}_{\delta\beta}.
\end{eqnarray*}
When the procedure is continued {\it ad-infinitum}, we find,
\begin{eqnarray*}
	G_{\alpha\beta} &=& G_{\alpha\alpha'}\delta_{\alpha'\beta}
	+ G_{\alpha\alpha'}M^o_{\alpha'\beta'}
	\widetilde{G}^{(\alpha)}_{\beta'\beta}
\\&&
	+ G_{\alpha\alpha'}M^o_{\alpha\delta}
	  \widetilde{G}_{\delta\delta'}^{(\alpha)} 
	  M^o_{\delta'\beta'}
	  \widetilde{G}^{(\alpha,\delta)}_{\beta'\beta}
\\
&&
	+ G_{\alpha\alpha'}M^o_{\alpha\delta}
	  \widetilde{G}_{\delta\delta'}^{(\alpha)} M^o_{\delta'\gamma}
	  \widetilde{G}^{(\alpha,\delta)}_{\gamma\gamma'}
	  M^o_{\gamma'\beta'}
	  \widetilde{G}^{(\alpha,\delta,\gamma)}_{\beta'\beta}
\\
&&
+\ldots
\end{eqnarray*}
This expression resembles that for $G_{\alpha\alpha'}$ in
\eq{expansion}, with $\beta=\alpha '$ and $\hat G_{\gamma\gamma'}$ replaced by
$\widetilde{G}_{\gamma\gamma'}^{(\alpha,\ldots)}$.  Furthermore, the
definition of $\widetilde{G}_{\gamma\gamma'}^{(\alpha,\ldots)}$ also
resembles the definition of $G_{\gamma\gamma'}$, but now with
restrictions on all wave-vector sets.  In fact the wave-vector
restrictions can be relaxed in the thermodynamic limit since the
restrictions remove only one term out of the sum over intermediate
wave-vector.  Relatively speaking, the difference between the series
with restricted and unrestricted sums is of order $O(1/N)$, so the
restriction on the intermediate wave-vectors is negligible in the
thermodynamic limit, and we can write
\[
	\widetilde{G}^{(\alpha,\ldots)}_{\beta\beta'}
	= G^{(\alpha,\ldots)}_{\beta\beta'},
\]
where $G^{(\alpha,\ldots)}_{\beta\beta'}$ is the full correlation
function which is diagonal in wave-vector {\it and} in which the
set of wave-vectors in $\beta$ differs from those in sets in $\alpha ,
\ldots$.

Thus \eq{expansion}, which has the non-physical $\hat G$, can be
replaced by the expansion
\begin{eqnarray}
	G_{\alpha\beta} &=& 
	G_{\alpha\alpha'}\delta_{\alpha' \beta}
	+ G_{\alpha\alpha'}M^o_{\alpha'\beta'}
	G^{(\alpha)}_{\beta'\beta}
\nonumber\\&&
	+ G_{\alpha\alpha'}M^o_{\alpha\delta}
	  G_{\delta\delta'}^{(\alpha)} 
	  M^o_{\delta'\beta'}
	  G^{(\alpha,\delta)}_{\beta'\beta}
\nonumber\\
&&
	+ G_{\alpha\alpha'}M^o_{\alpha\delta}
	  G_{\delta\delta'}^{(\alpha)} M^o_{\delta'\gamma}
	  G^{(\alpha,\delta)}_{\gamma\gamma'}
	  M^o_{\gamma'\beta'}
	  G^{(\alpha,\delta,\gamma)}_{\beta'\beta}
\nonumber\\
&&
+ \ldots ,
\eql{physicalexpansion}
\end{eqnarray}
which involves the full correlation function $G_{\alpha\alpha'}$ which
is diagonal in wave-vector.  In this expression none of the
intermediate wave-vector sets are allowed to be equal.
%The great appeal of the
%re-summation procedure is that now $G_{\alpha\alpha'}$, in
%turn, can be rewritten in terms of $G_{11}$ using \eq{Galphafactor}.

Using \eq{physicalexpansion}, we can write to leading $N$-order
\begin{equation}
	G_{12}(z) = G_{11}(z)*M_{12}(z)*G_{22}(z),
\eql{aG12}
\end{equation}
where, from \eq{Galphafactor}, $G_{22}(t)$ is given by, 
\begin{eqnarray}
	G^{22}_{\vek-\veqp,\veqp;\vek-\veq,\veq}
	(t)
	&\approx&
	G^{11}_{\vek-\veq}(t)
	\bullet
	G^{11}_{\veq}(t)\,
	\delta_{\veq,\vek-\veqp}
\nonumber\\&&	+
	G^{11}_{\vek-\veq}(t)\circ
	G^{11}_{\veq}(t)\,
	\delta_{\veq\veqp}
,
\eql{factorG22}
\end{eqnarray}
where superscripts like $22$ and $11$ are a reminder of the mode
orders of the arguments.  In order to facilitate writing tensor
products, we have introduced the following notational symbols for
products of tensors of rank $2$ ($A^{a;b}$), rank $3$ ($A^{a;bc}$ if
$b$ and $c$ belong to the same set $\alpha$ or $A^{ab;c}$ if $a$ and
$b$ belong to the same set), or rank $4$ (such as $A^{a,b;c,d}$):
\begin{eqnarray}
(A\cdot B)^{a;c} & \equiv &
	A^{a;b} B^{b;c} 
\nonumber \\
%(A \cdot B^T)^{a;c} & \equiv &
%	A^{b;a} B^{b;c} 
%\nonumber \\
(A : B )^{a;c} & \equiv & 
	A^{a;d,f} B^{d,f;c} 
\nonumber\\
(\inversion A)^{a b; c} &\equiv&
A^{c; b a} 
\nonumber \\
%(A : B^T )^{a;c} & \equiv & 
%	A^{a;d,f} B^{c;d,f} 
%\nonumber \\
(A \circ B)^{a,b;c,d} &\equiv&
	A^{a;c} B^{b;d}
\nonumber \\
(A \bullet B)^{a,b;c,d} &\equiv&
	A^{a;d} B^{b;c} ,
\eql{notation}
\end{eqnarray}
where repeated labels are summed over.

The two terms in the expression above for $G_{22}$ turn out to yield
the same contribution to $G_{12}$ in \eq{aG12}. However, this
summation has a pre-factor of $1/2$ from the number of ways the
indices can be interchanged in the ``$*$'' product.  Using
\eq{factorG22} in \eq{aG12}, we obtain 
%the expression for $G_{12}$,
\begin{eqnarray}
	G^{12}_{\vek;\vek-\veq,\veq}
	 (t)  
	&=& 
	\int_0^t
	G^{11}_{\vek}
	(t-\tau)
	\cdot
	\barM^{12}_{\vek;\vek-\veq,\veq}
\nonumber
\\&&
	\quad:\,
	[G^{11}_{\vek-\veq}(\tau)
	\circ
	G^{11}_{\veq}(\tau)] d\tau,
\eql{G12}
\end{eqnarray}
where the time convolution arises from the inverse Laplace transform. 

At this point, the necessity of including multi-linear modes is
readily apparent. For example, using the definition of the
multi-linear basis set, we can write
\begin{eqnarray}
	\saverage{Q_1(t)Q_1^*Q_1^*}
	&=& 
	\saverage{Q_1(t)Q_1^*}*K_{11}^{-1}*
	\saverage{Q_1Q_1^*Q_1^*}
\nonumber \\&&
	+G_{12}(t) * K_{22}\eql{C21} ,
\end{eqnarray}
and note that the second term would have been absent if the bi-linear
modes $Q_2$ in the basis set hadn't been included even though the
$N$-ordering of this term is the same as the first term.  Another
interesting point is that the first term is not present if the
subtractions in the definition of $Q_2$ in the basis set are not
included. If one assumes that the $Q_{1}$ are Gaussian random
variables, then correlations of the form $\saverage{Q_1Q_1^*Q_1^*}$
vanish.  However, in dense fluids the linear densities clearly do not
obey Gaussian statistics since static correlations such as
$\saverage{Q_1Q_1^*Q_1^*}$
% or the third virial coefficient 
involve configurational averages over the triplet distribution
function and are not negligible.  Note that the second term involves a
time convolution in \eq{G12} and can be expected to have quite
different behavior from the first term, which is proportional to an
ordinary time correlation function.

\subsection{Renormalization of the propagator}

\label{sec:renorm}

In this section we focus on the linear correlation function $G_{11}$
itself.  \eq{expansion} for $G_{11}(z)$ reads:
\begin{eqnarray*}
	G_{11} &=&
	\hatG_{11} +
	\hatG_{11} M^o_{1\alpha} \hatG_{\alpha\alpha'}
	 M^o_{\alpha'1}\hatG_{11} 
\\
	&& + \hatG_{11} M^o_{1\alpha} \hatG_{\alpha\alpha'}
	 M^o_{\alpha'\beta'} \hatG_{\beta'\beta}
	M^o_{\beta1} \hatG_{11}
\\&&
	+ \hatG_{11} M^o_{1\alpha} \hatG_{\alpha\alpha'}
	M^o_{\alpha'\beta'} \hatG_{\beta'\beta}
	M^o_{\beta\gamma} \hatG_{\gamma\gamma'} M^o_{\gamma'1} \hatG_{11}
\\&&
+ \ldots
\end{eqnarray*}
where for brevity we omitted the $z$ argument. In the summations over
$\alpha$, one can isolate all the terms with mode order $1$ to obtain,
\begin{eqnarray*}
	G_{11} &=& \hatG_{11} 
	+
	\hatG_{11} \big(M^o_{1\alpha}
	\hatG_{\alpha\alpha'} M^o_{\alpha'1}
	\big)\hatG_{11}
\\
	& &
	+ 
	\hatG_{11} \big(M^o_{1\alpha} \hatG_{\alpha\alpha'}
	M^o_{\alpha'\beta'} \hatG_{\beta'\beta} 
	M^o_{\beta1} \big)\hatG_{11}
\\&&
	+ 
	\hatG_{11} 
	\big(M^o_{1\alpha}
	\hatG_{\alpha\alpha'} M^o_{\alpha'\beta'} 
	\hatG_{\beta'\beta} M^o_{\beta\gamma}
	\hatG_{\gamma\gamma'} M^o_{\gamma'1} \big)\hatG_{11}
\\&	&+ 
	\hatG_{11} 
	\big (M^o_{1\alpha} \hatG_{\alpha\alpha'} M^o_{\alpha'1}\big)
	\hatG_{11}
	\big( M^o_{1\gamma} \hatG_{\gamma\gamma'} M^o_{\gamma'1}\big) 
	\hatG_{11}
\\&&
+ \ldots ,
\end{eqnarray*}	
where the summation over repeated indices here start at mode order 2.
Rearranging the terms, this can be written as
\begin{eqnarray}
	G_{11} &=& \hatG_{11} + \hatG_{11} * \Sigma_{11} * \hatG_{11}  
\nonumber
\\&&
		+ \hatG_{11} * \Sigma_{11} * \hatG_{11} *
		 \Sigma_{11} * \hatG_{11}  + \ldots,
\eql{G11}
\end{eqnarray}
with
\begin{eqnarray}
	\Sigma_{11} &=& \sum_{i=1}^{\infty} \Theta_i,
\nonumber\\
	\Theta_1 & = &\sum_{|\alpha|\neq1}M^0_{1\alpha} *
	\hatG_{\alpha\alpha'}*
	M^0_{\alpha'1}
\eql{Sigma}
\\ 
	\Theta_2 & = &
	\sum_{|\alpha|\neq1,|\beta|\neq1}
	M^0_{1\alpha}*  \hatG_{\alpha\alpha'} *
	M^0_{\alpha'\beta'} *\hatG_{\beta'\beta}* M^0_{\beta1}
\nonumber
\\	&&\ldots\nonumber
\end{eqnarray}
\eq{G11} can be re-summed as
\begin{equation}
	G_{11}(z) = [z{\identity} - M_{11} - \Sigma_{11}(z)]^{-1}
\eql{G11ren}
\end{equation}
where the inverse is taken on the $11$ sub-block level.  \eq{G11ren}
can be utilized to extract the complicated long-time dependence that
arises in the memory functions for the generalized Langevin when only
{\it linear} densities are included in the projection $\cal{P}$.  In
the theory of liquids, the Laplace transform of these memory functions
are generalized transport coefficients which reduce to the Green-Kubo
expressions in the limit of small $z$ and $k$.  \eq{G11ren} can be
cast in the form in which the full generalized transport coefficients
are expressed as a sum of bare transport coefficients and the
$\Sigma_{11}(z)$ terms, which renormalize the bare coefficients and
account for the complicated memory effects observed in dense
liquids\cite{SLO92}.  In the limit in which the energy density is
neglected in the basis set and only bilinear modes are included in the
multi-linear basis set, the idealized and extended mode coupling
theory models\cite{Gotze87} of the glass transition can be
obtained\cite{SLO92,Liu97}.

Finally, we mention that in Ref.~\cite{SLO92} it was shown that by
rearranging the terms, along the lines of Sec.\ref{subsec:multipoint},
$\hatG_{\alpha\alpha'}$ in the $\Theta_i$ can be replaced by the real diagonal
in wave-vector $G_{\alpha\alpha'}$, with the restriction in
summations over $\alpha$, $\beta$ etc, that none of their the
wave-vector set are identical. This diagonal $G_{\alpha\alpha'}$ will
factor again as in \eq{Galphafactor}. Then, \eq{G11ren}, \eq{Sigma}
and \eq{Galphafactor} lead to a self-consistent equation for $G_{11}$.

\section{Multi-time correlations}

\subsection{Separating slow and fast behavior in the context of multiple times}
\label{subsec:fast}

We now turn to multiple time correlation functions like
$\saverage{Q(t_2+t_1)Q(t_1)Q}$.  At first glance, it appears
attractive to use the formal formal solution of \eq{qdottwo},
\begin{eqnarray}
	Q_\alpha(t) = G_{\alpha\delta}(t)  Q_\delta(0) 
		+ \int_0^{t} G_{\alpha\delta}(t-\tau) 
		 \phi_\delta(\tau)d\tau,
\eql{formal}
\end{eqnarray}
to get an explicit expression for
$\saverage{Q_\alpha(t_1+t_2)Q_\beta(t_1)Q_\gamma^*}$. After all,
inserting \eq{formal} yields
\begin{eqnarray}
&&	\saverage{Q_\alpha(t_1+t_2)Q_\beta(t_1)Q_\gamma^*}		
\nonumber\\
&&	= \int_0^{t_1+t_2}\!d\tau_2 \int_0^{t_1}\!\!d\tau_1
	\, G_{\alpha\eta}(t_1+t_2-\tau_2)G_{\beta\delta}(t_1-\tau_1)
\eql{inprinciple}
\nonumber\\
	&&\times 
	\bigg[ \saverage{Q_\eta Q_\delta Q_\gamma^*}4\delta(\tau_2)\delta(\tau_1)
	+ \saverage{\phi_\eta(\tau_2)\phi_\delta(\tau_1)Q_\gamma^*} \bigg].
\end{eqnarray}
It is tempting to assume that the expression in \eq{inprinciple}
involving the fluctuating force behaves as
$\delta(\tau_2)\delta(\tau_1)$ at long times in an analogous fashion
to \eq{fastness1}, leading directly to a local equation in time,
\begin{equation}
   \saverage{Q_\alpha(t_1+t_2)Q_\beta(t_1)Q_\gamma^*}	
	\stackrel{?}{=} G_{\alpha\eta}(t_1+t_2)G_{\beta\delta}(t_1)
\widetilde{M}_{\eta\delta\gamma},
\eql{notright}
\end{equation}
where $\widetilde{M}_{\eta\delta\gamma}$ is related to the infinite
time integral of $[\saverage{Q_\delta Q_\delta Q_\gamma^*}
4\delta(\tau_2)\delta(\tau_1) +
\saverage{\phi_\eta(\tau_2)\phi_\delta(\tau_1)Q_\gamma^*}]$.  However,
things are not this simple as can be seen by using time translation
invariance to write
$\saverage{Q_\alpha(t_1+t_2)Q_\beta(t_1)Q_\gamma^*} =
\saverage{Q_\alpha(t_2)Q_\beta Q^*_\gamma(-t_1)}$. Applying again
\eq{formal}, we obtain
\begin{eqnarray*}
&&	\saverage{Q_\alpha(t_2)Q_\beta Q_\gamma^*(-t_1)}		
\nonumber\\
&&	= \int_0^{t_2}\!\!\!\!\!d\tau_2 \int_0^{-t_1}\!\!d\tau_1
	\, G_{\alpha\eta}(t_1+t_2-\tau_2)G^*_{\gamma\delta}(-t_1-\tau_1)
\nonumber\\
	&&\times 
	\bigg[ 
	\saverage{Q_\eta Q_\beta Q_\delta^*}4\delta(\tau_2)\delta(\tau_1)
	+ \saverage{\phi_\eta(\tau_2)Q_\beta \phi_\delta^*(\tau_1)} \bigg],
\end{eqnarray*}
of which the local time version would be
\begin{equation}
   \saverage{Q_\alpha(t_2)Q_\beta Q_{\hat\gamma}^*(-t_1)}
	K_{\hat\gamma\gamma} \stackrel{?}{=} G_{\alpha\eta}(t_2)
	\bar{M}_{\eta\beta\delta} G_{\delta\gamma}(t_1) \eql{right}
\end{equation}
where we have used the time-translation property,
$G_{\alpha\hat\beta}(t)K_{\hat\beta\beta} =
G^*_{\beta\hat\alpha}(-t)K_{\hat\alpha\alpha}$ (see Appendix
\ref{symmetry}), and defined
\begin{eqnarray}
	\bar{M}_{\delta\beta\theta} &=&
	\biggl\{
	-\int_0^\infty \!\!
	\int_0^\infty 
	\saverage{\phi_\delta(\tau_2) Q_\beta
	\phi^*_{\hat{\theta}}(-\tau_1)}
	d\tau_1d\tau_2
\nonumber
\\&&
	+\saverage{Q_\delta Q_\beta Q_{\hat\theta}^*}
	\biggr\} K^{-1}_{\hat{\theta}\theta}
\eql{firstM2}
\end{eqnarray}
Clearly, \eq{right} and \eq{notright} are in contradiction since, in
general, they will have different time behavior. 

The question of which (if either) instantaneous form is approximately
correct can be resolved by noting that products of the fluctuating
force $\phi(\tau)$ cannot always be treated as ``fast'', nor as
Gaussian random variables. This observation has been noted previously
by Schramm and Oppenheim\cite{SO81} who considered the quantity
$\saverage{\phi(\tau_1)\phi(\tau_2)\phi(\tau_3)}$ and showed that it
does not have a purely fast decay.  To understand this point, consider
the {\it equal} time correlation function
$\saverage{Q_\alpha(t)Q_\beta(t)Q_\gamma(t)} = \saverage{Q_\alpha
Q_\beta Q_\gamma}$. Inserting \eq{formal}, taking the limit
$t\rightarrow\infty$, and noting that in that limit $G(t)\rightarrow
0$, one obtains
\begin{eqnarray}
	\saverage{Q_\alpha Q_\beta Q_\gamma}
	&=&
	\lim_{t\rightarrow\infty}
	\int_0^t
	\int_0^t 
	\int_0^t  
	G_{\alpha\eta}(t-\tau_1)
	G_{\beta\delta}(t-\tau_2)
\nonumber
\\&&\times
	G_{\gamma\zeta}(t-\tau_3)
	\saverage{\phi_\eta(\tau_1)\phi_\delta(\tau_2)\phi_\zeta(\tau_3)}
\nonumber
\\&&
	\,d\tau_1\,d\tau_2\,d\tau_3.
\eql{nongaussian}
\end{eqnarray}
If $\phi(\tau)$ were Gaussian with zero mean, the three point
correlation on the right hand side would be zero, but since the left hand
side of \eq{nongaussian} doesn't vanish, $\phi(\tau)$ is not a Gaussian fluctuating
force. 

The three point correlation function in \eq{nongaussian} can't have a
purely fast decay either, since, by isotropy,
$\saverage{\phi_\eta(\tau_1)
\phi_\delta(\tau_2) \phi_\zeta(\tau_3)} = O(k^4)$, whereas the left
hand side is $O(k^0)$. It therefore follows that upon integration of the slow part of
$\saverage{\phi_\eta(\tau_1) \phi_\delta(\tau_2) \phi_\zeta(\tau_3)}$,
one needs to generate a factor $k^{-4}$. Schramm and Oppenheim
obtained the explicit form of the slow behavior of the three time
correlation function of the fluctuating force for the case of a single
slow variable $A$. It was shown that for $\tau_2$, $\tau_3 > \tau_1$
and small $|\vec{k}_1|$ (other cases are similar),
\begin{eqnarray*}
&&	\saverage{\phi_{\vec k_1}(\tau_1)
	\phi_{\vec k_2}(\tau_2)
	\phi_{\vec k_3}(\tau_3)
	}
\\
	&&\approx 
	2 |{\vec{k}}_1|^2 D e ^{-D|{\vec{k}}_1|^2|\tau_2-\tau_1|}
	\saverage{\hatA_{\vec k_1}\phi_{{\vec{k}}_2}(\tau_2-\tau_3)
	\phi_{{\vec{k}}_3}}
\end{eqnarray*}
where $D$ is the diffusion constant, which is clearly a
slowly-decaying function of $\tau_2-\tau_1$. On the other hand, it
appears that there is a fast decay in $\tau_2-\tau_3$, so that the
whole expression is small when {\em that} time difference becomes
large.  Nonetheless, the slow behavior in $\tau_2-\tau_1$ will bring
about a factor of $O(1/k^2)$ upon integration. For the full
restoration of the $O(k^0)$ term on the left hand side of
\eq{nongaussian}, we refer the reader to the original
paper\cite{SO81}.

Apparently, the assumption that $\cal{P}$ projects out all the slow
behavior is not sufficient to specify when a correlation function is
fast-decaying since clearly $\phi_\alpha(t)$ itself is not a fast
variable in every context.  Notice that despite the appearance of slow
behavior in the correlation function $\saverage{\phi_{\vec
k_1}(\tau_1)\phi_{\vec k_2}(\tau_2)\phi_{\vec k_3}(\tau_3)}$, there
are instances in which a multi-time correlation function of
fluctuating forces is certainly small; namely, when at least two of
the time arguments of the forces are well-separated. If the time
arguments of the fluctuating forces are not well-separated, slow
behavior can occur. In \eq{inprinciple}, we cannot assume that the
integrand is peaked around $(\tau_2,\tau_1)=(0,0)$ because these two
variables can come arbitrarily close in the integration over $\tau_2$
and $\tau_1$ along the line $\tau_1\approx\tau_2$ and slow behavior
can be expected. We conclude that the local time dependence of the
instantaneous \eq{notright} cannot be justified.

We propose the following general rule to determine when a correlation function is
fast decaying: {\em in a correlation function involving fluctuating
forces,} {\em the function decays quickly in a pair of time arguments,}
{\em provided these are well-separated in time}.  Note that ``well-separated'' here means
that the time difference is larger than the microscopic time
$\tau_m$. In applying this rule to situations when integrations are
carried out over the time arguments of fluctuating forces, 
we require that the time arguments can only get
close at isolated points which give contributions of measure zero to
the integral.

With these rules in mind, consider the correlation function
\[
	\saverage{\phi_\alpha(-\tau_2)\phi_\beta\phi_\gamma(\tau_1)}.
\]
This is clearly fast in $\tau_2$ as well as in $\tau_1$ provided $\tau_2$ and $\tau_1$
are positive. Therefore, \eq{right}
should be correct, as the correlations of the fluctuating forces in
\eq{firstM2} are of the form above, which we can write as
%re-write as
\[
\saverage{\phi_\alpha(-\tau_2)\phi_\beta\phi_\gamma(\tau_1)}=
	\average{\dot Q_\alpha e^{\calfastL \tau_2}\calP_\perp
	\phi_\beta e^{\calfastL \tau_1}\calP_\perp \dot Q}.
\]
We observe that the correlation function
above is fast in $\tau_{1}$ and $\tau_{2}$ provided $\tau_{1}\gg
\tau_{m}$ and $\tau_2\gg \tau_{m}$ since it has the form of a
succession of
two fast forward propagations which yields an expression
in which all time arguments are well-separated. 
If one of the times were negative, i.e.,
one of the propagators propagated backward in time, the expression
would no longer be (purely) fast. Hence an alternative way of
identifying terms which are fast in all time arguments is to require
that they have only forward fast propagation when applied in succession.

\subsection{Correlation functions involving multiple times}

Using the conclusions of the previous section, we will now derive a
recursion relation for multi-time correlation functions. We consider
the general case, denoted by
$G^{(n)}_{\alpha_n,\alpha_{n-1},\ldots,\alpha_0}(t_n,t_{n-1},\ldots,
t_{1})$ or $G^{(n)}_{\{\alpha_i\}}(\{t_i\})$, defined as 
\[
 G^{(n)}_{\{\alpha_i\}}(\{t_i\}) = 
\average{
Q_{\hat\alpha_0}^* Q_{\alpha_n}(t_1+\cdots+t_n) \ldots
Q_{\alpha_1}(t_1)} K^{-1}_{\hat{\alpha}_0\alpha_0}
,
\]
with all with $i=1\ldots n$ and $t_i\geq0$ so the arguments are
time-ordered. We write
%\begin{eqnarray*}
%&&
\begin{gather*}
\saverage{Q_{\hat{\alpha}_0}^*Q_{\alpha_n}(t_1+t_2+\cdots+t_n)\ldots
Q_{\alpha_2}(t_1+t_2)Q_{\alpha_1}(t_1)}
\\
%&&
\begin{align*}
\quad&=
	\average{Q_{\hat{\alpha}_0}^* e^{\calL t_1} Q_{\alpha_1}
	e^{\calL t_2} Q_{\alpha_2}
	\cdots
	e^{\calL t_{n-1}} Q_{\alpha_{n-1}}
	e^{\calL t_n} Q_{\alpha_n}}
\\
%&&
	&\equiv
	\saverage{X e^{\calL t_{n-1}} B}.
\end{align*}
\end{gather*}
%\end{eqnarray*}
where $B= Q_{\alpha_{n-1}}Q_{\alpha_n}(t_n)$ and $X$ is the  operator
\begin{eqnarray*}
X &=& Q_{\hat{\alpha}_0}^* e^{\calL t_1} Q_{\alpha_1}
	e^{\calL t_2} Q_{\alpha_2}
	\cdots
	e^{\calL t_{n-1}}.
\end{eqnarray*}

Using \eq{eLtexp} applied to $B$, and inserting the result into
$\saverage{Xe^{\calL t_{n-1}}B}$, we get
%\begin{eqnarray}
\begin{multline}
\average{X e^{\calL t_{n-1}} B } 
%&=&
=
%\\
%&&
\saverage{Q_\delta^*\,B} 
K^{-1}_{\delta\hat\delta}  \saverage{X Q_{\hat\delta}(t_{n-1}) }
%\nonumber
\\
%&&
        - \int_0^{t_{n-1}} \!\!
	\saverage{\dot{Q}_\delta^*\, e^{\calfastL\tau_1}\calP_\perp
	 B} 
%\nonumber
%\\
%&&
%\times
	 K^{-1}_{\delta\hat\delta}  \saverage{X
Q_{\hat\delta}(t_{n-1}-\tau_1)}
 d\tau_1 ,
\\
+\saverage{X e^{\calfastL t_{n-1}}
\calP_\perp B}
%\nonumber
\eql{firstQQQ}
\end{multline}
%\end{eqnarray}
From the discussion in Sec.~\ref{subsec:fast}, it is now apparent that
the third term can be considered fast in $t_{n-1}$ because only
forward propagation occurs in $X$ since all $t_i$ are positive, and there is a
projected propagation in $t_{n-1}$.  For macroscopic times for which
$t_{n-1} \gg \tau_{m}$, this term can be neglected. Note, however,
this term could not be be neglected in integrals of \eq{firstQQQ} over
the time $t_{n-1}$.  
Inserting $B=Q_{\alpha_{n-1}}Q_{\alpha_n}(t_n)$ and using \eq{formal},
one obtains
\begin{eqnarray}
G^{(n)}_{\{\alpha_i\}}(\{t_i\})
% &=&  G_{\alpha_n\beta}(t_n)\saverage{Q_\beta Q_{\alpha_{n-1}} Q^*_\delta}
%	K_{\delta\hat\delta}^{-1}
%\nonumber\\
%&&\times 
%	G^{(n-1)}_{\hat\delta,\alpha_{n-2},\ldots}(t_{n-1},\ldots)
%\nonumber
%\\
%&&
% -   \int_0^{t_n-1}\!\!\!d\tau_1\int_0^{t_n}\!\!d\tau\,
%	G_{\alpha_n\beta}(t_n-\tau)
%\nonumber
%\\
%&&
% \times
%	\saverage{Q^*_\delta\,e^{\calP_\perp\calL \tau_1}\calP_\perp
%	Q_{\alpha_{n-1}}\phi_\beta(\tau)}
%	K_{\delta\hat\delta}^{-1} 
%\nonumber
%\\
%&&
%	\times
%	G^{(n-1)}_{\hat\delta,\alpha_{n-2},\ldots}(t_{n-1}-\tau_1,t_{n-2},\ldots)
%\nonumber\\&&
% + \fastterm{t_{n-1}}
%\nonumber
%\\
&=&
 \int_0^{t_{n-1}}\!\!\!\int_0^{t_n}\!\!\!\,
	G_{\alpha_n\beta}(t_n-\tau)
	M_{\beta\alpha_{n-1}\delta}(\tau,\tau_1)
\nonumber
\\
&&
	\times
	G^{(n-1)}_{\delta,\alpha_{n-2}\ldots}(t_{n-1}-\tau_1,t_{n-2},\ldots) \;
	 d\tau d\tau_1
\nonumber\\&&
 + \fastterm{t_{n-1}}
\eql{thirdQQQ}
\end{eqnarray}
where
\begin{eqnarray}
	M_{\beta\alpha\delta}(\tau,\tau_1) &=& \bigg(
	4\average{Q_\beta Q_\alpha
	Q_{\hat\delta}}\delta(\tau)\delta(\tau_1)
\nonumber\\
&&
	-\average{
	\dot{Q}^*_{\hat{\delta}}
	e^{\calfastL\tau_1}\calP_\perp
	\phi_\beta(\tau) Q_\alpha }
	\bigg) K^{-1}_{\hat{\delta}\delta}
\eql{M2}.
\end{eqnarray}
\eq{thirdQQQ} is the desired recursion relation. Neglecting the term
which is fast in $t_{n-1}$, the $n$ multi-time correlation functions
can be related to the $n-1$ multi-time correlation, and thus
ultimately in terms of $G^{(1)}_{\alpha_1\alpha_0}(t_1)
=G_{\alpha_1\alpha_0}(t_1)$.  The instantaneous version of
\eq{thirdQQQ} is simply
\begin{equation}
	G^{(n)}_{\alpha_n,\ldots}
	(t_n,\ldots)
=
  G_{\alpha_n\beta}(t_n) 
	\barM_{\beta \alpha_{n-1}\delta}  
	G^{(n-1)}_{\delta,\alpha_{n-2},\ldots}(t_{n-1},\ldots)
\eql{instantmultiQ},
\end{equation}
where $\barM_{\delta 1\theta}=
\int_0^\infty d\tau_1 \int_0^\infty d\tau
M_{\delta 1\theta}(\tau,\tau_1)$. 

Applying \eq{instantmultiQ} to the three time correlation function 
$G_{\alpha\gamma\beta}(t_2,t_1)$, \eq{right} is recovered:  
\begin{equation}
  G_{\alpha \gamma\beta}(t_2,t_1)
=G_{\alpha\delta}(t_2) 
\barM_{\delta \gamma\theta}  G_{\theta\beta}(t_1)
\eql{instantQQQ}.
\end{equation}  
Again using the recursion relation, one can also derive
equations for correlation functions involving four or more
times, e.g., 
\begin{equation}
G^{(3)}_{\alpha\beta\gamma\delta}(t_3,t_2,t_1) =
	G_{\alpha\zeta}(t_3)\barM_{\zeta\beta\theta}
	G_{\theta\eta}(t_2)\barM_{\eta\gamma\lambda}
	G_{\lambda\delta}(t_1).
\end{equation}
It is clear now that any multi-time correlation $G^{(n)}$ can be
written as a product of $n$ factors of $G_{11}$ and $n-1$ vertices
$\barM$.

\subsection{$N$-ordering of triple time correlation function}

In this section, we consider the leading $N$-order and mode
coupling terms expressions for the triple time
correlation function for linear densities in which
$|\alpha|=|\beta|=|\gamma|=1$.  Taking \eq{instantQQQ} for the linear
densities, this correlation function is given by 
\begin{equation}
  G_{\alpha 1\beta}(t_2,t_1) =G_{\alpha\delta}(t_2)
\barM_{\delta 1\theta}  G_{\theta\beta}(t_1)
\eql{instantAAA}.
\end{equation}
From this relation, it is evident that
the $N$-ordering of $\barM_{\gamma 1\delta}$ follows from that of 
$G^{-1}_{\gamma\alpha}(t_2) G_{\alpha1\beta}(t_2,t_1)
G_{\beta\delta}^{-1}(t_1)$, and hence we need
to establish the $N$-ordering properties of $\average{Q_\alpha(t_2+t_1)Q_1(t_1)Q_\beta}$. 
%We start
%by considering a few examples, 
%\begin{gather*}
%	\average{Q_1(t_2+t_1)Q_1(t_1)Q_1} =  O(N) \\
%	\average{Q_2(t_2+t_1)Q_1(t_1)Q_1} =  O(N^2) \\
%	\average{Q_2(t_2+t_1)Q_1(t_1)Q_2} =  O(N^2) \\
%	\average{Q_3(t_2+t_1)Q_1(t_1)Q_1} =  O(N^2) \\
%	\average{Q_3(t_2+t_1)Q_1(t_1)Q_2} =  O(N^3) 
%	\average{Q_3(t_2+t_1)Q_1(t_1)Q_3} =  O(N^3) 
%	\average{Q_4(t_2+t_1)Q_1(t_1)Q_1} =  O(N^2) \\
%	\average{Q_4(t_2+t_1)Q_1(t_1)Q_2} =  O(N^3) \\
%	\average{Q_4(t_2+t_1)Q_1(t_1)Q_3} =  O(N^4) \\
%	\average{Q_4(t_2+t_1)Q_1(t_1)Q_4} =  O(N^4) 
%\end{gather*}
%This makes us suspect that
One can show, by induction in $|\beta|$, that
\[
	\average{Q_\alpha(t_2+t_1)Q_1(t_1)Q_\beta}  = 
	\left\{\begin{array}{ll}
		O(N^{|\alpha|+1}) 
			& \mbox{if $|\alpha|<|\beta|$}\\
		O(N^{|\beta|}) 
			& \mbox{if $|\alpha|=|\beta|$}\\
		O(N^{|\beta|+1}) 
			& \mbox{if $|\alpha|>|\beta|$}\\
	\end{array}
	\right.
.
\]
%This is easy to prove for $|\beta|=1$ and can be proved inductively
%for the rest.
Combining with \eq{thirdQQQ} and \eq{NorderKinv}, we obtain
\begin{equation}
	G_{\alpha1\beta}(t_2,t_1)  =  
	\left\{\begin{array}{ll}
		O(N^{1-(|\beta|-|\alpha|)}) 
			& \mbox{if $|\alpha|<|\beta|$}\\
		O(N^0) 
			& \mbox{if $|\alpha|=|\beta|$}\\
		O(N^1) 
			& \mbox{if $|\alpha|>|\beta|$}
	\end{array}
	\right. .
\eql{N-ordering2}
\end{equation}
Using \eq{N-ordering1} and \eq{N-ordering2} to establish the $N$-order of 
$G^{-1}_{\gamma\alpha}(t_2)
G_{\alpha1\beta}(t_2,t_1) G_{\beta\delta}^{-1}(t_1)$, one finds that
the $N$-ordering of $\barM_{\gamma 1\delta}$ follows the same
$N$-ordering rules as $G_{\gamma1\delta}(t_2,t_1)$.

With the $N$-ordering expressions above, the dominant
contributions to $G_{111}(t_2,t_1)$ are given by
\begin{equation}
\begin{split}
	G_{111}(t_2,t_1)=& G_{11}(t_2) 
	* \barM_{111} * G_{11}(t_1)
\\+&		 G_{12}(t_2) * \barM_{211} * G_{11}(t_1)
\\+&		 G_{11}(t_2) * \barM_{112} * G_{21}(t_1)
+ O(N^{-1})
\eql{lead}
.
\end{split}
\end{equation}
The leading mode-coupling corrections to \eq{lead} involve a large
number of terms of order $N^{-1}$.  Collecting these
additional terms gives a net factor of $M \approx
V k_c^{3}$, so that the sum of all $N^{-1}$ terms gives a term of
order $ M/N \approx (k_c \xi)^{3}$ which survives in the thermodynamic limit.
To first order in $M/N$, one obtains the following correction terms:
\begin{multline*}
	G_{13}(t_2) * \barM_{312} * G_{21}(t_1)
	+ G_{13}(t_2) * \barM_{311} * G_{11}(t_1)
\\
	+ G_{12}(t_2) * \barM_{212} * G_{21}(t_1)
	+ G_{12}(t_2) * \barM_{213} * G_{31}(t_1)
\\
	+ G_{11}(t_2) * \barM_{113} * G_{31}(t_1)
\end{multline*}

These results can be easily extended to higher order correlations $G^{(n)}_{\{\alpha_i\}}$ with $n>3$
and where $|\alpha_i|=1$ since all the necessary $N$ orderings are known.
Thus, any multi-time correlation function of $Q_1$'s can 
be expressed in terms of the vertices $\barM_{\alpha 1 \beta}$ and {\it two}-time
(but possibly multi-point) correlation functions. In turn, the
multi-point correlation functions can be expressed in terms
of the vertices $\barM_{\alpha\beta}$ and the linear propagator
$G_{11}(t)$ as explained in Sec.~\ref{subsec:multipoint}, and hence
all time dependences in multi-time correlation functions
at long times can be expressed in terms of the two-time correlation
functions of linear densities.  From the mode-coupling formalism, the two-time
correlation functions can be evaluated self-consistently via the
relations \eq{G11ren} and \eq{Sigma}.

Using the results from Sec.~\ref{subsec:multipoint}, and \eq{lead} and
\eq{G12}, one can obtain the leading $N$-order expressions for $G_{111}$ in
terms of $\barM_{11}$, $\barM_{111}$ and $G_{11}(t)$ as follows.  Inserting
the reduced forms of the vertices $M_{211}$ and $M_{112}$ that are
derived in Appendix \ref{appendix} into \eq{lead} yields,
\begin{eqnarray}
  G^{111}_{\vek-\veq,\veq,\vek}(t_2,t_1)
 = H_1 +H_2 +H_3,
\eql{ourthreetime}
\end{eqnarray}
where
\begin{eqnarray*}
   H_1 &=&	G^{11}_{\vek-\veq}(t_2)
		\cdot
		\bar{M}^{111}_{\vek-\veq;\veq;\vek}
		\cdot
		G^{11}_{\vek}(t_1)
\\
  H_2 &=&
	G^{12}_{\vek-\veq;-\veq,\vek}(t_2)
	: \Big[
	K^{11}_{\veq}
	\circ
	G^{11}_{\vek}(t_1)
	\Big]
\\
  H_3 &= &
	G^{11}_{\vek-\veq}(t_2)
	\cdot
	G^{21}_{\vek-\veq,\veq;\vek}
	(t_1)
\end{eqnarray*}
In $H_2$, we can use \eq{G12} to express
$G^{12}_{\vek-\veq;-\veq,\vek}(t_2)$ in terms of the linear
$G_{11}(t)$ and vertices $\barM$. For $G_{21}$ in $H_3$, using the
fact that $G_{\alpha\beta}(t) = G_{\hat\beta\hat\alpha}(t)
K_{\hat\alpha\alpha}K^{-1}_{\hat\beta\beta}$ [\eq{B3}],
% and that complex
%conjugation inverts the wave-vectors, 
an analogous expression
for $G_{21}$ is obtained,
\bleq
\begin{eqnarray*}
	G^{21}_{\vek-\veq, \veq;\vek}
	 (t_1)  
	&=& \inversion
	\int_0^{t_1}
	K^{-1}_{\vek}\cdot G^{11}_{\vek}
	(\tau_1)
	\cdot
	\barM^{12}_{\vek;\vek-\veq,\veq}
	\,:\,
	\Big\{
	\big[G^{11}_{\veq}(t_1-\tau_1) \cdot K_{\veq}\big]
	\circ
	\big[G^{11}_{\vek-\veq}(t_1-\tau_1)\cdot 
	K_{\vek-\veq}\big]
	\Big\} 
	d\tau_1.
\end{eqnarray*}
%\begin{eqnarray*}
%	G_{\vec k_1-\vec k_2,\vec k_2;\vec k_1}(t_1)
%	&=&
%	\int_0^{t_1} 
%	\Big\{
%	\big[
%	 G_{\vec k_1-\vec k_2}(t_1-\tau_1)
%	 \cdot
%	K_{\vec k_1-\vec k_2}
%	\big]
%\\&&	\quad
%	\circ
%	\big[ 
%	 G_{\vec k_2}(t_1-\tau_1)
%	 \cdot
%	 K_{\vec k_2} 
%	\big]
%	\Big\} 
%\\&&\quad
%	: 
%	\barM^{*T}_{\vec k_1;\vec k_1-\vec k_2,\vec k_2}
%	\cdot
%	\big[
%	K^{-1}_{\vec k_1}
%	\cdot
%	G_{\vec k_1}(\tau_1) 
%	\big]^T
%d\tau_1
%\end{eqnarray*}
%This can also be derived using that $G_{21}=G_{22}: M_{21}
%\cdot G_{11}$ and $\barM_{21} = K_{22}: \big[
%K^{-1}_{11} \cdot \barM^{*}_{12} \big]^T$.  
Thus, in
\eq{ourthreetime}, we have
\begin{eqnarray}
	H_2 &=& 
	\int_0^{t_2}
	G^{11}_{\vek-\veq}(\tau_2)
	\cdot
	\barM^{12}_{\vek-\veq;-\veq,\vek}
	:
%\nonumber\\&&\quad
	\Big\{ \big[ G^{11}_{-\veq}(t_2-\tau_2)\cdot K_{\veq}
	       \big]
%\nonumber\\&&\qquad
	\circ \;
	\big[ G^{11}_{\vek}(t_2-\tau_2)\cdot G^{11}_{\vek}(t_1)
	\big] \Big\}
d\tau_2,
\eql{48}
\\
	H_3 &=&
%	\int_0^{t_1}
%	\Big\{ \big[ G_{\vec{k}_1-\vec{k}_2}(t_2) 
%	 \cdot \; G_{\vec{k}_1-\vec{k}_2}(t_1-\tau_1) 
%	 \cdot  K_{\vec{k}_1-\vec{k}_2} \big]
%%\nonumber\\&&\qquad
%	\circ \; \big[
%	 G_{\vec{k}_2}(t_1-\tau_1)
%	\cdot \;
%	 K_{\vec{k}_2}
%	\big] \Big\}
%%\nonumber\\&&\qquad
%	:
%	  \barM^{*T}_{\vec{k}_1;\vec{k}_1-\vec{k}_2,\vec{k}_2}
%	\cdot
%	\big[
%	K^{-1}_{\vec{k}_1}
%	\cdot
%	G_{\vec{k}_1}(\tau_1)
%	\big]^T
%d\tau_1.
	\inversion
	\int_0^{t_1}
	K^{-1}_{\vek}\cdot G^{11}_{\vek}
	(\tau_1)
	\cdot
	\barM^{12}_{\vek;\vek-\veq,\veq}
	\,:\,
	\Big\{
	\big[G^{11}_{\veq}(t_1-\tau_1) \cdot K_{\veq}\big]
	\circ
	\big[G^{11}_{\vec{k}_1-\vec{k}_2}(t_1-\tau_1)\cdot 
	G^{11}_{\vek-\veq}(t_2)\cdot
	K_{\vek-\veq}\big]
	\Big\} 
	d\tau_1.
\eql{49}
\end{eqnarray}
In a subsequent publication\cite{nextArticle}, we demonstrate that the
expressions above give excellent results for a moderately
dense hard-sphere system in the hydrodynamic regime.

\eleq

\section{Comparison with Kawasaki's theory: non-Gaussian effects}
Roughly $20$ years ago, Ronis \cite{R81} examined higher-order
correlation functions within the Kawasaki mode-coupling
formalism\cite{K70,K77}. The treatment itself is too technical to
recapture here, so we will simply state the results from that paper to
compare with those from the present theory.  For a multi-point
correlation, Ronis obtains [his Eq.~(3.11)]
\begin{eqnarray}
C_{k_0k_1k_2}^{\alpha_0\alpha_1\alpha_2}(t)
&\equiv&
\cumulant{A_{\alpha_0,k_0}(t)A_{\beta_1,-k_1}A_{\beta_2,-k_2}}
\nonumber\\&&
\times \saverage{A_{k_1}A^*_{k_1}}^{-1}_{\beta_1\alpha_1}
\saverage{A_{k_2}A^*_{k_2}}^{-1}_{\beta_2\alpha_2}
\nonumber\\
&=&
2 \int_0^t dt_1 G_{\alpha_0\beta_0}(k_0,t-t_1) 
V^{\beta_0\beta_1\beta_2}_{k_0k_1k_2}
\nonumber\\&&
\times G_{\beta_1\alpha_1}(\vec k_1,t_1)G_{\beta_2\alpha_2}(\vec{k}_2,t_1)
\eql{RonisC21}
\end{eqnarray}
We keep Ronis's notation here, as it is close enough to ours to be
understood.  \eq{RonisC21} looks essentially like \eq{G12}, but is missing
the first term in \eq{C21} as might be expected from a Gaussian
theory for three-point correlation functions.  Intriguingly, this first term
is all one would obtain for the three-point correlation function 
from a projection operator approach if only
linear densities were included in the basis set for the long-time
dynamics. In addition, the vertex $V$ in the Gaussian theory also 
differs from the vertex $M_{21}$ due to the subtraction
terms in the basis set which are not present.  These differences in
functional form of the vertices can significantly alter the time
profile of both multi-point and multi-time correlation functions\cite{nextArticle}.

In Ref.~\cite{R81}, the following expression for a three-time correlation
function was derived [Eq.~(6.5b) therein],
\begin{gather*}
\cumulant{A_{\alpha_0,k_0}(t_0)A_{\beta_1,k_1}^*(t_1)A_{\beta_2,k_2}^*(0)}
\saverage{A_{k_1}A^*_{k_1}}^{-1}_{\beta_1\alpha_1}
\nonumber
\\
\begin{align}
\qquad=& 2 \int_0^{t_0} d\tau_1 G_{\alpha_0\beta_0}(k_0,t_0-\tau_1)
	   V^{\beta_0\beta_1\beta_2}_{k_0,-k_1,k_2}
\nonumber\\
&\times
	G_{\beta_2\alpha_2}(k_2,\tau_1)
	\saverage{A_{\beta_1,k_1}(\tau_1)A_{\alpha_1,k_1}(t_1)}
\nonumber\\
\qquad+&2 \int_0^{t_1} d\tau_1 G_{\alpha_1\beta_0}(k_1,t_1-\tau_1)
	 V^{\beta_0\beta_1\beta_2}_{k_1,-k_0,k_2}
\nonumber\\
&
\times
	G_{\beta_2\alpha_2}(k_2,\tau_1)
	\saverage{A_{\beta_1,-k_0}(\tau_1)A_{\alpha_0,k_0}(t_0)}.
\eql{starr}
\end{align}
\nonumber
\end{gather*}

Comparison between \eq{starr} and the expression for the three-time
correlation function in \eq{ourthreetime} is facilitated by noting that
in the instantaneous approximation and neglecting mode-coupling
corrections, we can write
\begin{eqnarray*}
	G^{11}_{\vek}(t_2-\tau_2)
	\cdot
	G^{11}_{\vek}(t_1)
	&=& 
	G^{11}_{\vek}(t_1+t_2-\tau_2);
\nonumber\\
	G^{11}_{\vek-\veq}(t_2)
	\cdot
	G^{11}_{\vek-\veq}(t_1-\tau_1) 
&=& 
	G^{11}_{\vek-\veq}(t_1+t_2-\tau_1).
\end{eqnarray*}
From careful inspection of \eq{starr} and \eq{ourthreetime}, one sees
that $H_{3}$ (see \eq{49}) is essentially equivalent to the second
term \eq{starr}.  However the term $H_{2}$ differs from the first term
in \eq{starr} in two ways.  First, the way in which the indices are
contracted with the vertex $V^{\beta_0\beta_1\beta_2}_{k_0,-k_1,k_2}$,
as written in Ref.~\cite{R81}, differs from the tensor contractions in
$H_{2}$.  Second, and more importantly,
there seem to be significant differences in the upper limits in
the time-convolution integrals, which in Ref.~\cite{R81} is $t_0=t_1+t_2$, as
opposed to $t_2$.  This is particularly intriguing in light of the
observation that the upper limit of $t_0$ was obtained in
\eq{notright} where the time dependence of the correlation function of
the fluctuating forces was treated incorrectly.
Nonetheless, in both mode-coupling theories, the higher order time
correlation functions are expressed in terms of ordinary (two) time
correlation functions.  The major differences between the theories arise because of the Gaussian
approximation in the Kawasaki formalism. Hence it is not
surprising that some (static) three point correlations are
missed since they vanish if the linear densities $Q_1$ are assumed to obey
Gaussian statistics at all times.  This deficiency was noted by Ronis
who suggested that these differences result in significant deviations
only at short times.  It is clear from the present formalism, however,
that this is not the case since terms of 
the form $G_{11}(t_2) * \barM_{111} * G_{11}(t_1)$ decay slowly in both
$t_{1}$ and $t_{2}$.  These findings have been confirmed in numerical
simulations of hard sphere systems\cite{nextArticle}.

\section{Higher-order correlation functions for the ideal gas}
\label{sec:ideal}

In this section, the mode coupling formalism will be illustrated for an
ideal gas system in the grand canonical ensemble composed of particles
of mass $m$ in a volume $V$ at an inverse temperature $\beta$. In the
ideal gas, the motion of each particle $j$ is given by
\[
	r_j(t) = r_j(0) + \frac{p_j(0)}{m} t
;\quad
	p_j(t) = p_j(0).
\]
Given the simple form of the particles trajectories, 
any time correlation function can be calculated exactly and
compared with the expressions that follow from mode coupling
theory.

\subsection{Conserved quantities}
An essential step in applying the formalism to a particular system is
the identification of the slow variables of the system.
In any gas composed of point particles, particle number, momentum and
energy are conserved and hence their corresponding
densities, 
%$N_k$, $\vec{P}_k$, and $E_k$ 
are slowly-varying quantities. 
The ideal gas system is quite different from simple gases in that it has many more conserved
quantities since the momentum $p_{j}$ of each particle is conserved
along all directions.  Consequently, a tagged particle density 
of the form $p_{jx}^{m_x} p_{jy}^{m_y}p_{jz}^{m_z}
e^{i{\vec{k}}\cdot\vec p_j}$, where $m_x$, $m_y$ and $m_z$  are 
arbitrary integers, should be included for each particle $j$ in the set of slow variables.
However, for collective modes, it is not hard to show that it suffices
to include densities of all analytical functions of the momenta
$f(\vec{p}_{j})$, i.e.,
\[
	\sum_{j=1}^N f(\vec{p}_{j}) e^{i{\vec{k}}\cdot\vec r_j},
\]
since the contribution of a single tagged particle to correlation of
extensive variables is $O(1/N)$.

Taking the Hermite polynomials $H_n$ as a basis for the
functions $f(p)$, where
\[
	H_0(u) = 1; 
	H_1(u) = 2u; 
	H_n(u)=(-)^ne^{u^2}\ddh{^n}{u^n}e^{-u^2},
\]
the complete set of linear slow variables is given by
\begin{eqnarray}
	A^{\{i\}}_{{\vec{k}}} &=& \sum_{j=1}^N 
	\frac{H_{i_x}(u_j^x)H_{i_y}(u_j^y)H_{i_z}(u_j^z)}
	{\sqrt{2^{i_x+i_y+i_z}i_x!\,i_y!\,i_z!}}
	 e^{i\vec k\cdot\vec r_j},
\eql{idealslowset}
\end{eqnarray}
where $\{i\}$ denotes the set of three indices $\{i_x, i_y, i_z\}$, each
of which runs from zero to infinity.  If we define $u^x_j=p_j^x\sqrt{\beta/2m}$,
$u^y_j=p_j^y\sqrt{\beta/m}$ and $u^z_j=p_j^z\sqrt{\beta/m}$, then the
inner product of the Hermite polynomials corresponds to the canonical average,  
\begin{eqnarray}
\int_{-\infty}^{\infty} H_n(u) H_m(u) \frac{e^{-u^2}}{\sqrt\pi} du.
	&=& 2^n n!\delta_{mn} 		\eql{Hnormalization} \\
 &=& \saverage{ H_{n}(u) H_{m}(u) } \nonumber
.
\end{eqnarray}
Since $\saverage{H^i(u)}=0$ unless $i=0$, $\hat
A^{\{i\}}_{\vec{k}}$ is given by
\[
	\hat A^{\{i\}}_{{\vec{k}}} =  A^{\{i\}}_{{\vec{k}}}
- \saverage{N} \delta_{i_x0}\delta_{i_y0}\delta_{i_z0}\delta_{{\vec{k}} 0},
\]
and the correlation function
$\saverage{\hatA^{\{i\}}_{{\vec{k}}}\doubles{\hatA^{\{j\}}_{{\vec{k}}}}^*}$
is given by
\[
\saverage{\hatA^{\{i\}}_{{\vec{k}}}\doubles{\hatA^{\{j\}}_{{\vec{k}}}}^*}=
\saverage{N}\delta_{i_xj_x}\delta_{i_yj_y}\delta_{i_zj_z}.
\]

\subsection{Two-time, two-point correlation functions}

Consider the density mode $\hat N_k$, which corresponds to
$A^{\{0\}}_{{\vec{k}}}$ ($\{0\}$ is short for $\{0,0,0\}$). For
${\vec{k}}\neq0$, the density-density time correlation function is
given by
\begin{eqnarray*}
	G\Qarg{\{0\}\{0\}}{{\vec{k}}}(t) 
%&=& 
%	\frac{ \saverage{\hat{N}_{k}(t) \hat{N}_{k}^{*}}}
%	{\saverage{\hat{N}_{k} \hat{N}_{k}^{*}}}\\ 
	&=& 
\frac{\baverage{\sum_{j=1}^N\sum_{l=1}^N
	e^{i{\vec{k}}\cdot(\vecr_j(t)-\vecr_l)}}
	}{
%	\baverage{\sum_{j=1}^N\sum_{l=1}^Ne^{i{\vec{k}}\cdot(\vecr_j-\vecr_l)}}
\saverage{N}
	}
 = \saverage{e^{i{\vec{k}}\cdot\vecp_1t/m}}.
\end{eqnarray*}
This result follows from the fact that
$\saverage{\sum_{i=1}^N\exp(\vecr_i\cdot{\vec{k}})}=\saverage{N}\delta_{{\vec{k}}0}$,
and the statistical independence of particles.  As the momentum $\vecp_1$
is Gaussian-distributed, one obtains
\begin{eqnarray}
   G\Qarg{\{0\}\{0\}}{{\vec{k}}}(t) &=& \int_{-\infty}^{\infty}
	\frac{e^{i|{\vec{k}}|pt/m}e^{-\beta p^2 /(2m)}}{\sqrt{2\pi
	m/\beta}} dp 
\nonumber
\\&=& e^{-|{\vec{k}}|^2t^2 / 2m\beta}
	= e^{-(\tilde{k}t)^2/2} ,
	\eql{G11exact}
\end{eqnarray}
where $\tilde{k}=|\vec k|/\sqrt{m\beta}$ is a conveniently scaled
wave-vector.  

We will compare the exact result \eq{G11exact} to the result from the
mode-coupling framework of this paper obtained using
Eqs.~(\ref{eq:qdottwo}) and (\ref{eq:M}).  The first point to note is
that $\dot{Q}_{\alpha}(t)$ is proportional to $Q_{\alpha}$, which
follows from the facts the all Hermite polynomials have been included
in the set of slow variables and that $(d/dt)H_{n}(u)=0$ for all
$n$.  This, in turn, implies that $\calP_\perp\dot{Q}_\alpha=0$, and
hence the fluctuating force $\phi_{\alpha}(t)$ vanishes for all
$t$. According to Eqs.~(\ref{eq:M}) and (\ref{eq:instantM}),
$M_{\alpha\beta}(\tau) = 2\delta(\tau)\bar{M}_{\alpha\beta}$, since
$\phi_{\alpha}(t)=0$.  Thus \eq{qdotthree} is exact, with
\begin{equation}
	\bar{M}_{\alpha\beta} 
	= \saverage{\dot{Q}_\alpha Q_{\hat\beta}^*} 
	* K_{\hat\beta\beta}^{-1} .
\eql{idealM}
\end{equation}

Since $\dot{Q_1}$ is can be written as a linear combination of $Q_1$'s,
and, in general, $dot{Q}_\alpha = \sum_{|\beta|\leq|\alpha|}a_{\alpha\beta}Q_\beta$,
we conclude that $\saverage{\dot{Q}_\alpha
Q_\beta^*}=0$ for $\beta>\alpha$ since the multi-linear basis set is
orthogonal in mode-order by construction.  Similarly, since
$\saverage{\dot{Q}_\alpha Q_\beta} = -
\saverage{Q_\alpha\dot{Q}_\beta^*}$, $\saverage{\dot{Q}_\alpha
Q_\beta}=0$ for $\alpha > \beta$ and hence 
$\bar{M}_{\alpha\beta}$ is diagonal in mode order, implying that  \eq{qdotthree}
decouples at each mode order $n$ into equations for the multi-point correlation functions $G_{nn}(t)$.

Focusing on the linear variables, \eq{qdotthree} reduces to
\[
	\dot{G}\Qarg{\{i\}\{j\}}{{\vec{k}}}(t) = 
	\sum_{\{l\}}
	\bar{M}\Qarg{\{i\}\{l\}}{{\vec{k}}} G\Qarg{\{l\}\{j\}}{{\vec{k}}}(t)
\]
Due to the orthogonality of the $A^{\{i\}}_{{\vec{k}}}$'s, 
$G\Qarg{\{i\}\{j\}}{{\vec{k}}}(0)=\delta_{\{i\}\{j\}}$,
and hence,
\begin{equation}
	G\Qarg{\{i\}\{j\}}{{\vec{k}}}(t) 
	= \left[\exp\bigl(\bar{M}\Qarg{11}{{\vec{k}}}t\bigr)\right]
	_{\{i\}\{j\}} .
\eql{Gidealprin}
\end{equation}

The structure of the matrix $\bar M_{11}$ is relatively simple, which makes
it possible to actually calculate $G_{11}(t)$ from
\eq{Gidealprin}. Fixing the direction of $\vec k$ to be along $\hat
x$, for $|{\vec{k}}|\neq0$ we note that,
\begin{eqnarray*}
	\dot{A}{}_{{\vec{k}}}^{\{j\}}
%	&=& 
%	\frac{ik_x}{\sqrt{m\beta/2}}
%	\sum_{n=1}^N 
%	\frac{u_n^xH_{j_x}(u_n^x) H_{j_y}(u_n^y)H_{j_z}(u_n^z)}
%	{\sqrt{2^{j_x+j_y+j_z}j_x!\,j_y!\,j_z!}}
%	 e^{ik_xr^x_n}
%\\
&=&
	i\tilde{k}
%	\frac{ik_x}{\sqrt{m\beta}} 
	\left[
	\sqrt{j_x+1}\,
	\hatA_{{\vec{k}}}^{\{j_x+1,j_y,j_z\}}
	+\sqrt{j_x}\,
	\hatA_{{\vec{k}}}^{\{j_x-1,j_y,j_z\}}
	\right] ,
\end{eqnarray*}
where Eqs.~(\ref{eq:idealslowset}) and the recursion relation for the
Hermite polynomials
\[
	2u H_{n-1}-2(n-1)H_{n-2} =H_n(u),
\]
have been used.  With \eq{idealM}, this leads to the tri-diagonal form
for the matrix $\barM_{11}$ at linear order,
\begin{eqnarray}
	\barM\Qarg{\{i\}\{j\}}{{\vec{k}}} 
	&=& 
%\frac{ik_x}{\sqrt{m\beta}}
	i\tilde{k}
	\delta_{i_yj_y}\delta_{i_zj_z}
\nonumber\\&&
	\times
	(\sqrt{i_x+1}\,\delta_{i_x+1,j_x} + \sqrt{i_x}
	\,\delta_{i_x-1,j_x}).
\eql{Mtridiagonal}
\end{eqnarray}
The terms in \eq{Mtridiagonal} can be conveniently expressed as the expectation values of
the raising and lowering operators in the Hermite basis representation
of the quantum Harmonic oscillator.
Identifying
\[
	\ket n   = \hatA^{\{n,j_y,j_z\}}_{{\vec{k}}}
\]
and defining
\[
	\bracket{m}{\calB\,n} \equiv 
	\saverage{\doubles{\hatA^{\{m,j_y,j_z\}}_{{\vec{k}}}}^*
	\calB\hatA^{\{n,j_y,j_z\}}_{{\vec{k}}} }
	/ \saverage{N}
\]
for any operator $\calB$, and 
\[
	a \ket n  = \sqrt{n} \ket{n - 1};	
\quad
	a^\dagger \ket n = \sqrt{n+1} \ket{n + 1},
\]
one has
\[
	\bracket{m}{a\, n} = \sqrt{n}\,\delta_{m,n-1}
;\quad
	\bracket{m}{a^\dagger \, n} = \sqrt{n+1}\,\delta_{m,n+1}.
\]
In this representation, the matrix $\barM \Qarg{\{i\}\{j\}}{{\vec{k}}}$
can be written as
\[
\barM \Qarg{\{i\}\{j\}}{{\vec{k}}}=\bracket{j_x}{\bar\calM\,i_x} ,
\]
where $\calM$ is the operator
\begin{equation}
	\bar\calM = i\tilde{k}
%	\frac{ik_x}{\sqrt{m\beta}} 
(a^\dagger+a)
.
\eql{Mop}
\end{equation}
In essence, $\bar\calM$ is nothing but 
the Liouville operator $\cal L$,
restricted to act on the space of phase space functions which are
linear combinations of $\hatA_{\vec{k}}$.

To reproduce \eq{G11exact}, according to  \eq{Gidealprin}, we must evaluate
\[
	\bbracket{0}{e^{\bar\calM t} 0} ,
\]
which can be done in straightforward fashion using the 
Campbell-Baker-Hausdorff formula\cite{anybook}.
This formula states that if $\calA$ and $\calB$ are
linear operators, a linear operator $\calC$ exists such that
$e^{\calA} e^{\calB} = e^{\calC}$, where $\calC$
of the form $\calC = \calA + \calB +
\sfrac12 [\calA,\calB]$ plus repeated commutators.
Taking $\calA = i\tilde{k}t a^{\dagger}$ and $\calB = 
i\tilde{k} ta$ and noting that $[a,a^\dagger] = 1$, \eq{Mop} yields,
\[
e^{i\tilde kt a^\dagger}
    e^{i\tilde kt a} 
=    e^{\bar\calM t}e^{(\tilde kt)^2/2} ,
\]
which can be re-arranged to give,
\begin{equation}
	e^{\bar\calM t} = e^{i\tilde kt a^\dagger}
	    e^{i\tilde kt a} 
		e^{-(\tilde{k}t)^2/2} ,
\eql{Moprelation}
\end{equation}
and therefore
\begin{eqnarray*}
	\bbracket{0}{e^{\bar\calM t}0}	
%	&=& \bbracket{0}{e^{i\tilde kt a^\dagger} 
% 	e^{i\tilde k t a} \, 0 } e^{-(\tilde kt)^2/2}
%\\
&=& \bbracket{ e^{-i\tilde k t a} 0 }{
 e^{i\tilde kt a} 0} 
 e^{-(\tilde{k}t)^2/2}.
\end{eqnarray*}
But as $e^{i\tilde k ta} \ket 0 = \ket 0$, 
\[
	\bbracket{0}{e^{\bar\calM t} 0}
	=
	e^{-(\tilde kt)^2/2}.
\]
which coincides, as expected, with the exact result \eq{G11exact}.

The time correlations for other basis functions are now
easily obtained using \eq{Moprelation}. For example, 
\begin{eqnarray}
	\frac{\baverage{A^{\{j\}}_{\vec{k}}(t) \doubles{A^{\{0,0,0\}}_{\vec{k}}}^*}
	}{\saverage{N}}
	&=& \bbracket{0}{ 
	    e^{i\tilde kt a} 
		j_x}
		e^{-(\tilde k t)^2/2}
	\delta_{j_y0}\delta_{j_z0}
\nonumber
\\
	&=&
	\frac{(it\tilde k)^{j_x}}{\sqrt{j_x!}}
	e^{-(\tilde kt)^2/2}
	\delta_{j_y0}\delta_{j_z0}.
\eql{j0}
\end{eqnarray}

Note that it is only with an infinite set of conserved quantities
that it is possible for a system with no dissipative terms to lead to
relaxation. One might argue that taking a finite number of slow
variables, 
%say $N_{\vec{k}}$, $\vec P_{\vec{k}}$ and $E_{\vec{k}}$, 
(say, mass, momentum and energy density)
the
dissipative terms in $\barM$ would no longer be zero. While this is correct
and $\barM$ is not purely imaginary, $\barM$ would still
be finite in each mode block, and one would get exponential decay
in the instantaneous limit instead of Gaussian decay for all correlation functions.

\subsection{Two-time, multi-point correlation functions}

Consider the correlation function 
\[
   C^{\{0\}\{0\}\{0\}}
	_{\vek-\veq,\veq;\vek}
	(t)\equiv
	\saverage{\hat N_{\vek-\veq}(t)\hat N_{\veq}(t)
	\hat N^{*}_{\vek}}
/ \saverage{N},
\]
with $|\vek|\neq0$, $|\veq|\neq0$ and $\vek\neq\veq$.  The direct
calculation of this correlation function is simple if one notes that
\begin{eqnarray*}
\baverage{\sum_{l=1}^N 
e^{i(\vek-\veq)\cdot\vec r_l(t)}
\sum_{m=1}^N e^{i{\veq}\cdot\vec r_m(t)}
\sum_{n=1}^N e^{-i{\vek}\cdot\vec r_n} }
/ \saverage{N}
\\
= \baverage{\sum_{l=1}^N e^{i(\vek-\veq)\cdot\vec r_l(t)}
e^{i\veq\cdot\vec r_l(t)}
e^{-i\vek\cdot\vec r_l} },
\end{eqnarray*}
since any term in the summation over particle indices in which $l \neq
m$, $l \neq n$ or $m \neq n$ yields a Kronecker delta for one of the
wave-vectors.  Such terms do not give a contribution if all
wave-vectors are non-zero. The wave-vector $\veq$ drops out of the
expression, and
\begin{eqnarray}
   C^{\{0\}\{0\}\{0\}}
	_{\vek-\veq,\veq;\vek}
	(t)
	&=&
 \baverage{\sum_{l=1}^N 
e^{i\vek\cdot\vec p_lt/m}}
/\saverage{N}
\nonumber\\
	&=& G\Qarg{\{0\}\{0\}}{\vek}(t) 
	= e^{-(\tilde k t)^2/2}.
\eql{Q21exactideal}
\end{eqnarray}

The mode coupling derivation of this same result goes as follows.  
From \eq{G12}, noticing that $M_{21}=0$, we immediately obtain $G_{21}(t)=0$.  
The simplest argument for the vanishing of $G_{21}(t)$ is that since
$\barM$ is diagonal in mode order, no correlations between different
mode orders exist, so that
\[
	G_{\alpha\beta}(t) = 0 \quad\mbox{if $|\alpha|\neq|\beta|$}.
\]

From the definition of the multi-linear basis set in \eq{Qset}, 
$C^{\{0\}\{0\}\{0\}}_{\vek-\veq;\veq;\vek}(t)$ can be written
in terms of the linear and bi-linear densities as
\begin{eqnarray*}
	 C^{\{0\}\{0\}\{0\}}
	_{\vek-\veq,\veq;\vek}(t)
	&=& 
	\saverage{Q^{\{0\}\{0\}}
	_{\vek-\veq,\veq}(t)\hat N^{*}_{\vek}}
/
\saverage{N}\\
&&
+
\saverage{\hat N_{\vek-\veq}
	\hat N_{\veq}\hat N^{*}_{\vek}}
\saverage{\hat N_{\vek}(t)\hat N^{*}_{\vek}}
/\saverage{N}^2. 
\end{eqnarray*}
However since $G_{21}(t)=0$, $C^{\{0\}\{0\}\{0\}}
_{\vek-\veq,\veq;\vek}(t)$ gets its value solely
from the subtraction terms. Furthermore, since $\saverage{\hat
N_{\vek-\veq}\hat N_{\veq}\hat N^{*}_{\vek}}=\saverage{N}$ for an ideal gas system, one gets
\begin{eqnarray*}
	 C\QQzQargg{\{0\}}{\vek-\veq}{\{0\}}{\veq}{\{0\}}{\vek}(t)
	&=& 
\saverage{\hat N_{\vek}(t)\hat N^{*}_{\vek}}
/\saverage{N},
\end{eqnarray*}
in agreement with \eq{Q21exactideal}.

\subsection{Multi-time correlation functions}

Finally, we conclude our discussion of the higher-order correlation
functions in the ideal gas by examining the validity of the
expressions for the multi-time correlation functions in
\eq{instantQQQ}.  For the three time correlation function for the
density mode $N_k$, direct calculation gives,
\begin{eqnarray}
	G\QQQargg{\{0\}}{\vek-\veq}{\{0\}}{\veq}{\{0\}}{\vek}
	(t_2,t_1)
	&=& \baverage{\sum_{l=1}^N 
	e^{i\vec p_l\cdot[\vek(t_1+t_2)-\veq t_2]}}
	/\saverage{N}
\nonumber\\
	&=& e^{-|\tilde\vek(t_1+t_2)-\tilde\veq t_2|^2/2}
\eql{exactG111}
\end{eqnarray}
where $\tilde\vek = \vek/\sqrt{m\beta}$ and
$\tilde\veq = \veq/\sqrt{m\beta}$.

If $\vek$ and $\veq$ are both along $\hat x$, the
mode-coupling expression for the multi-time correlation [see
\eq{instantQQQ}] is
\begin{eqnarray*}
	G\QQQargg{\{0\}}{\vek-\veq}{\{0\}}{\veq}{\{0\}}{\vek}
	(t_2,t_1)
	&=& 
%\sum_{\{i\}\{j\}}
	  G\Qarg{\{0\}\{j\}}{\vek-\veq}(t_2)
	  \barM^{ \{j\}\{0\}\{i\}}_{\vec{k-q},\vec{q},\vec{k}}
%\\
%&&\times
	  G\Qarg{\{i\}\{0\}}{\vek}(t_1),
\end{eqnarray*}
where the repeated sets of indices $\{i\}$ and $\{j\}$ are summed.
Since $G\Qarg{\{i\}N}{}(t)$ was previously evaluated in \eq{j0}, and 
$G\Qarg{\{i\}N}{}(t)=\doubles{G\Qarg{N\{i\}}{}}^*(-t)$, the only unknown
quantity in this expression is 
\[
	\barM\QQQargg{\{j\}}{\vek-\veq}{\{0\}}{\veq}{\{i\}}{\vek} 
	= \baverage{\hatA\Qarg{\{j\}}{\vek-\veq}
		\hatA\Qarg{\{0\}}{\veq}
		\doubles{\hatA\Qarg{\{i\}}{\vek}}^{*}
		}
	/\saverage{N}.
\]
Writing this out using \eq{idealslowset}, the only surviving terms in
the summations over particle index for an ideal gas system are the
terms where all indices are equal, so
\[
	\barM\QQQargg{\{j\}}{\vek-\veq}{\{0\}}{\veq}{\{i\}}{\vek} 
	= \delta_{\{i\}\{j\}}.
\]
Hence
\begin{eqnarray*}
G
\QQQargg{\{0\}}{\vek-\veq}
	{\{0\}}{\veq}
	{\{0\}}{\vek}
(t_2,t_1)
	&=&  
	\sum_{j_x=0}^{\infty}
	 e^{-|\tilde{\vek}-\tilde{\veq}|^2t_2^2/2}
	   e^{-|\tilde{\vek}t_1|^2/2}
\\
&&\times
 \frac{1}{j_x!}\big[
	-t_1t_2\tilde{k}_{x}(\tilde{k}_{x}-\tilde{q}_{x})
	\big]^{j_x}.
\end{eqnarray*}
This can be summed to 
\[
e^{-[|\tilde{\vek} -\tilde{\veq}|^2t_2^2
-|\tilde{\vek}|^2t_1^2
-\tilde{k}_{x}(\tilde{k}_{x}-\tilde{q}_{x})t_1t_2 ]/2},
\]
which corresponds to the exact result \eq{exactG111}.

\section{Summary}

In this paper, a mode coupling theory was presented in which
multi-point and multi-time correlation functions are expressed in
terms of ordinary two-point, two-time correlation functions and a set
of vertices. The mode coupling theory developed here does not assume
that 
%random 
fluctuating forces (noise) are Gaussian distributed and
therefore is a generalization of mode coupling theories based on
Kawasaki's formalism\cite{K70,K77}. Furthermore, unlike kinetic theories,
it is not restricted to low densities and should be applicable to
dense fluids where cooperative motions of particles and collective
modes are important.

The formalism is based on projection operator techniques, which, for
ordinary two-point, two-time correlation functions, lead to a
generalized Langevin equation in which the memory function decays on a
microscopic time scale.  The simple extension of the projection
operator formalism to multi-time correlation functions is complicated
by the fact that the fluctuating forces appearing in the generalized
Langevin equation do not obey Gaussian statistics.  Furthermore,
multiple time correlations of the fluctuating force can in fact have a
slow decay when the time arguments of these forces become comparable.

In order to treat multiple time correlation functions of 
%random
fluctuating forces properly, the correlation functions were massaged
so that the time arguments of all fluctuating forces appearing in the
correlations were guaranteed to be well-separated, ensuring that all
memory functions which arise in the mode coupling theory decay to zero
on a molecular time scale.  This construction allows equations which
are local in time to be obtained which relate the multi-time
correlation function to two-time but multi-point correlations coupled
by essentially time-independent vertices.  The multi-point correlations, in turn,
can be written as convolutions of two-point and two-time
correlation functions coupled by time-independent vertices.  These
correlation functions can either be taken directly from experiment, simulation,
or can be solved self-consistently within the mode coupling formalism.
The vertices, which are composed of a static part (Euler term) and a
generalized transport coefficient, can similarly be calculated from
kinetic theory or taken from molecular dynamics and Monte-Carlo
simulations.

The equations for higher-order correlation functions contain an
infinite sum of terms which can be made tractable for systems with a
finite correlation length by applying a
cumulant expansion technique, termed the $N$-ordering method.  The
method was applied to obtain the leading order and first mode coupling
corrections of expressions for multi-point and multi-time correlation functions.
A key step in the $N$-ordering method and the proper setup of the theory is the
definition of an {\em orthogonal} multi-linear basis. Although in
principle it is possible to apply cumulant expansion methods to other
choices of a multi-linear basis, the orthogonalization procedure
simplifies the perturbation analysis enormously and helps to avoid
erroneous truncations of the mode coupling series.

The expressions for the higher-order correlation functions bear a
resemblance to those found by Ronis\cite{R81} within the framework of
Kawasaki's mode coupling theory.  In this approach, the linear
densities composing the set of slow variables are assumed to be
Gaussian random variables at all times.  Although the Gaussian
assumption provides another method of simplifying the mode coupling
series for higher-order correlation functions, certain terms are
absent from the Gaussian theory which are neither small nor
quickly-decaying.  

The mode coupling predictions for higher-order correlation functions
for an ideal gas system were calculated analytically and shown to give
the exact results for both multi-time and multi-point correlation
functions.  An essential step in arriving at the correct result was the
inclusion of a complete set of densities in the
set of slow variables.  Although the ideal gas system does not
constitute a rigorous test of the formalism since all fluctuating
forces vanish, it is important to note that 
the formal mode coupling theory expressions for the higher-order
correlation functions yield the exact result.  In a future
publication\cite{nextArticle}, we compare the mode coupling
predictions for the multi-point and multi-time correlation functions
for a hard sphere fluid to data from simulations.  The theoretical
predictions of all higher-order correlation functions are in
remarkable agreement with the simulation results in the hydrodynamic
regime provided both Euler and dissipative vertex couplings are
included.

The theory outlined here has obvious applications to multi-dimensional
Raman and NMR spectroscopy, and simulation studies of dynamic
heterogeneity in dense fluids, glasses and polymers.  Since the theory
involves physical correlation functions, it is well-positioned to
address fundamental issues in characterizing the dynamics in systems
exhibiting non-exponential relaxation processes and frustration.  In
fact the current formalism has been used in Refs.\cite{SLO92,Liu97} to
justify some of the approximations made in mode-coupling theory for
the supercooled liquids\cite{Gotze75,Gotze87}.  These avenues are
currently being investigated.

\section*{Acknowledgments}

The authors would like to thank Irwin Oppenheim and David Reichman for
useful discussions.  This work was supported by a grant from the
Natural Sciences and Engineering Research Council of Canada and funds
from the Premier's Research Excellence Award.

\appendix

\section{Reduction of the vertices $\barM_{211}$ and $\barM_{112}$}
\label{appendix}

Neglecting mode coupling corrections, the vertices $\barM_{211}$ and
$\barM_{112}$ can be reduced to very simple forms.
The strategy used to simplify these terms is similar to that used for the factorization of
$\barM_{\alpha\beta}$ in Ref.~\cite{SLO92}. 
First $\barM_{112}$ is re-written with the help of \eq{instantQQQ} as,
\[
    \barM_{112} = G^{-1}_{1\alpha}(t_2)G_{\alpha1\beta}(t_2,t_1)
				G^{-1}_{\beta2}(t_1).
\]
Using the $N$-ordering of $G_{\alpha1\beta}(t_2,t_1)$, 
$G_{1\alpha}(t_2)$, and $G_{\beta 2}(t_1)$, one sees that to leading $N$-order,
\begin{equation}
\barM_{112} = G^{-1}_{11}(t_2)G_{112}(t_2,t_1)
				G^{-1}_{22}(t_1).
\eql{M112lead}
\end{equation}
In this expression, the leading $N$-order contributions are obtained
from the part of the various factors in \eq{M112lead} which are
diagonal in wave-vector.

By the property in \eq{factorization}, $K_{22}$ can be factored as,
\begin{eqnarray*}
&&\baverage{\hatA^a_{\vek-\veq}(t_1+t_2)
	  \hatA^b_{\veq}(t_1)
	   \doubles{Q\QQarg{c}{\vek-\veqp}{d}{\veqp}}^*}
\\
&&\approx
	\baverage{\hatA^a_{\vek-\veq}(t_1+t_2)
		  \hatA^{c^*}_{\vek-\veq}}
	\baverage{\hatA^b_{\veq}(t_1)
		  \hatA^{d*}_{\veq}}\delta_{\veq\veqp}
\\
&& 
	+ \baverage{\hatA^a_{\vek-\veq}(t_1+t_2)
		  \hatA^{d^*}_{\vek-\veq}}
	\baverage{\hatA^b_{\veq}(t_1)
		  \hatA^{c*}_{\veq}}\delta_{\vek-\veq,\veqp} .
\end{eqnarray*}
In a similar fashion, $K_{22}^{-1}$ factors to leading $N$-order,
\begin{eqnarray*}
K^{22^{-1}}_{\vek-\veq,\veq;\vek-\veqp,\veqp}
&\approx&
	K^{-1}_{\vek-\veq}
	\circ
	K^{-1}_{\veq}
	\delta_{\veq\veqp}
\\&&	+
	K^{-1}_{\vek-\veq}
	\bullet
	K^{-1}_{\veq}
	\delta_{\vek-\veq,\veqp},
%\\
%
%
%K\QQzQQarg{^{-1}\,c}{\vek-\veq}{d}{\veq}
%	{e}{\vek-\veqp}{f}{\veqp}
%&\approx&
%	K\Qarg{-1\,ce}{\vek-\veq}
%	K\Qarg{-1\,df}{\veq}
%	\delta_{\veq\veqp}
%\\&&	+
%	K\Qarg{-1\,cf}{\vek-\veq}
%	K\Qarg{-1\,de}{\veq}
%	\delta_{\vek-\veq,\veqp},
\end{eqnarray*}
where the notation introduced in \eq{notation} has been used.  This leads to the
following factorization of the wave-vector-diagonal part of
$G_{112}(t_2,t_1)$:
\begin{gather*}
\begin{align*}
&
%\qquad\qquad\approx
		G^{112}_{\vek-\veq;\veq;
	   \vek-\veqp,\veqp}
	(t_2,t_1) \approx G^{11}_{\vek-\veq}(t_1+t_2)\circ G^{11}_{\veq}(t_1)
	\delta_{\veq\veqp}
\\&\qquad\qquad
	+
	G^{11}_{\vek-\veq}(t_1+t_2)\bullet G^{11}_{\veq}(t_1)
	\delta_{\vek\veq-\veqp}.
\end{align*}
\end{gather*}
The factorization of $G_{22}^{-1}$ can be worked out as well:
\begin{eqnarray*}
	G^{22^{-1}}_{\vek-\veq,\veq;
		\vek-\veqp,\veqp}
	(t_1)
	&\approx&
	G^{11^{-1}}_{\vek-\veq}(t_1)\circ
	G^{11^{-1}}_{\veq}(t_1)
	\delta_{\veq\veqp}
\\&+&
	G^{11^{-1}}_{\vek-\veq}(t_1)
	\bullet
	G^{11^{-1}}_{\veq}(t_1)
	\delta_{\vek\veq-\veqp}.
\end{eqnarray*}
Inserting these expression into \eq{M112lead} and using \eq{instantG}, yields
\begin{eqnarray}
&&\barM^{112}_{\vek-\veq;\veq;
	\vek-\veqp,\veqp}
\nonumber\\
&&\approx 
	G^{11^{-1}}_{\vek-\veq}(t_2)\cdot
	G^{11}_{\vek-\veq}(t_2+t_1)
	\cdot
	G^{11^{-1}}_{\vek-\veq}(t_1)
	\circ {\bf 1}
	\delta_{\veq\veqp}
\nonumber\\
&&+
	G^{11^{-1}}_{\vek-\veq}(t_2)
	\cdot
	G^{11}_{\vek-\veq}(t_2+t_1)
	\cdot
	G^{11^{-1}}_{\vek-\veq}(t_1)
	\bullet
	{\bf 1}
	\delta_{\veq\veqp}
\nonumber\\
&&\approx {\bf 1}\circ{\bf 1}\delta_{\veq\veqp}
	+ {\bf 1}\bullet{\bf 1}\delta_{\vek-\veq,\veqp} ,
\eql{M112reduced}
\end{eqnarray}
where ${\bf 1}^{ac}=\delta_{ac}$.

Finally, \eq{M112reduced} and the relation $\barM_{211} =
K_{22}*(\barM_{1{1^{^*}}2})^* * K^{-1}_{11}$ give the leading
$N$-order expression for $M_{211}$,
\begin{eqnarray}
	\barM^{211}_{\vek-\veqp+\veq,	
			\veqp;-\veq;\vek}
\approx
	K_{\veq}
	\circ 
	{\bf 1}
	\,\delta_{\vek\veqp} 
%	K\Qarg{bd}{\veq}
%	\delta_{ac}\delta_{\veq\veqp} 
	+
%	K\Qarg{cd}{\veq}\delta_{ad}\delta_{\veq\veqp} 
	 {\bf 1}
	\bullet
	K_{\veq}
	\,\delta_{\veq\veqp} ,
\eql{M211reduced}
\end{eqnarray}
to lowest order in the mode-coupling parameter $M/N$.
It interesting that the first term in \eq{M2} gives the only
contribution to $M_{211}$ at order $(M/N)^{0}$: In other words, when
mode-coupling corrections are neglected, the term involving the 
fluctuating forces in \eq{M2} can be dropped.

Note that in this appendix, we have allowed ourselves to neglect
the difference between the inverse of quantities on the multi-linear
level and the linear-linear sub-block level, which is correct to order
$(M/N)^{0}$ and consistent with the level of approximation of
the rest of the derivation of \eq{M112reduced} and \eq{M211reduced}.

\newbox{\targ}
\savebox{\targ}{\mbox{$t$}}
\section{Symmetry properties of $G_{\alpha\beta}(\usebox{\targ})$ and 
$\bar{M}_{\alpha\beta}$.}
\label{symmetry}

In the ideal gas case, the $Q_\alpha$'s are either even or odd
functions of momentum. This property implies that the elements of
the multi-linear basis set $Q_{\alpha}$ are either symmetric or
anti-symmetric under the time-reversal operator ${\cal T}$, which
reverses the momenta.  Mathematically, this property can be written as
${\cal T} Q_\alpha = \gamma_{\alpha} Q_\alpha$, where
$\gamma_{\alpha}$ is either $1$ if $\alpha$ contains an even number of
momenta indices, or $-1$ when it has an odd number.  In addition, 
there typically is a symmetry under the reflection operator ${\cal R}$ which
inverts both the momenta and the positions of all particles. 
As the basis set elements $Q_\alpha$ depends on the spatial degrees of
freedom through $\exp (i\vec{k} \cdot \vec{r}_{j})$ [see \eq{idealslowset}], it
follows that ${\cal R}Q_\alpha = \gamma_\alpha Q_\alpha^*$.  
These relations also holds for other systems in which the potential
energy depends only on the distances between particles in the system.
These two symmetries plus time translation invariance have
the following implications for $\bar{M}_{\alpha\beta}$ and
$G_{\alpha\beta}(t)$ (see also Ref.~\cite{BP76}).

As the equilibrium distribution function is invariant under $\cal R$, 
\begin{eqnarray}
G_{\alpha\beta}(t) &=&\saverage{Q_\alpha(t)Q_{\hat\beta}^*}
	K_{\hat\beta\beta} 
= \saverage{({\cal R}Q_\alpha(t))({\cal R}Q_{\hat\beta}^*)}
	K_{\hat\beta\beta} 
\nonumber\\
&=&\gamma_\alpha\gamma_\beta G_{\alpha\beta}^*(t).
\eql{B1}
\end{eqnarray}
Hence, if $\gamma_\alpha\gamma_\beta=1$, the imaginary part is zero,
otherwise the real part is zero. 
Since the wave-vector dependence of the densities always enters in the form of $i$ times a
wavevector,  imaginary correlation functions must be odd functions of
the wave-vector and real correlation must be even functions of
wave-vector, provided these quantities are analytic in wave-vector.

Time reversal invariance, ${\cal T} e^{{\cal L}t} = e^{-{\cal
L}t}{\cal T}$, yields
\begin{eqnarray*}
	\baverage{\big(e^{{\cal L}t}Q_\alpha\big)Q_\beta^*} &=&
	\baverage{\big({\cal{T}}e^{-{\cal{L}}t}{\cal{T}}Q_\alpha\big)Q_\beta^*}
	= 
	\baverage{\big(e^{-{\cal{L}}t}{\cal{T}}Q_\alpha\big){\cal{T}}Q_\beta^*}
\\
	&=& \gamma_\alpha\gamma_\beta
	\baverage{\big(e^{-{\cal{L}}t}Q_\alpha\big)Q_\beta^*}
\end{eqnarray*}
and hence $G_{\alpha\beta}(t) = \gamma_\alpha\gamma_\beta
G_{\alpha\beta}(-t)$. This can be 
combined with Eq.~(\ref{eq:B1}) to yield
\begin{equation}
	G_{\alpha\beta}(t) = G_{\alpha\beta}^*(-t).
\eql{B2}	
\end{equation}

So from Eqs.~(\ref{eq:B1}) and (\ref{eq:B2}), we conclude that if
$\gamma_\alpha\gamma_\beta=1$, $G_{\alpha\beta}(t)$ is real, even in
wavevectors and symmetric under $t\rightarrow-t$, whereas if
$\gamma_\alpha\gamma_\beta=-1$, it is imaginary, odd in wavevectors and
anti-symmetric under time reversal. As $\barM=(dG(t)/dt) G^{-1}(t)$, this also implies
that $\barM_{\alpha\beta}$ is real and even in wavevectors
if $\gamma_\alpha\gamma_\beta=1$, and imaginary and odd in wavevector
otherwise.  

The following ordering is now valid when the magnitudes of the 
wave-vectors are small.  Imaginary correlation functions and vertices,
being odd in the wavevector arguments, are typically of linear order in
the wave-vectors.  But as the vertices $\bar M$ contains time
derivatives, they will always be at least of the order of the
wave-vectors, so a real-valued vertex is at least quadratic in
wave-vector, whereas a real-valued correlation function is typically
of order one.

Finally, using time-translation invariance,
%$\saverage{Q_\alpha(t)Q_\beta^*}=\saverage{Q_\alpha Q_\beta^*(-t)}$,
%or 
$G_{\alpha\beta}(t) = G^*_{\hat\beta\hat\alpha}(-t)K_{\hat\alpha\alpha}
K^{-1}_{\beta\hat\beta}$,
which conveniently combines with Eqs.~(\ref{eq:B2}) to 
\begin{equation}
	G_{\alpha\beta}(t) = G_{\hat\beta\hat\alpha}(t)
	K_{\hat\alpha\alpha}
	K^{-1}_{\beta\hat\beta}.
\eql{B3}
\end{equation}

\ecols

\end{document}